\documentstyle[12pt]{article}
\let\ssection=\section
\renewcommand{\section}{\setcounter{equation}{0}\ssection}

\newcommand\ba{\begin{array}}
\newcommand\ea{\end{array}}
\newcommand{\be}{\begin{enumerate}}
\newcommand{\ee}{\end{enumerate}}
\newcommand{\bi}{\begin{itemize}}
\newcommand{\ei}{\end{itemize}}
\newcommand{\bd}{\begin{description}}
\newcommand{\ed}{\end{description}}
\newcommand{\beq}{\begin{equation}}
\newcommand{\eeq}{\end{equation}}
\newcommand{\beqa}{\begin{eqnarray}}
\newcommand{\eeqa}{\end{eqnarray}}

\newcommand{\cf}{{\em cf\/}}

\newcommand{\etc}{{\em etc}}
\newcommand{\eq}[1]{(\ref{#1})}
\newcommand{\eqs}[2]{(\ref{#1}--\ref{#2})}
\newcommand{\ie}{{\em i.e.,\ }}

\renewcommand{\a}{\alpha}                   
\renewcommand{\b}{\beta}                    
\renewcommand{\d}{\delta}		    

\newcommand{\f}{\phi}
\newcommand{\g}{\gamma}

\renewcommand{\l}{\lambda}

\newcommand{\r}{\rho}
\newcommand{\s}{\sigma}
\newcommand{\th}{\theta}

\newcommand{\map}{\rightarrow}

\newcommand{\op}[1]{\widehat{#1}}

\newcommand{\tr}{{\rm tr}}

\newcommand\C{\mkern1mu\raise2.2pt\hbox{$\scriptscriptstyle|$}
                {\mkern-7mu\rm C}}
\newcommand{\R}{{\rm I\! R}}                

\newcommand{\U}{{\cal U}}
\newcommand{\UP}{{\cal U}{\cal P}}
\renewcommand{\S}{{\cal S}}
\renewcommand{\H}{{\cal H}}
\newcommand{\T}{{\cal T}}
\renewcommand{\L}{{\cal L}}
\newcommand{\M}{{\cal M}}
\renewcommand{\P}{\op{P}}
\newcommand{\Q}{\op{Q}}
\newcommand{\h}[1]{({#1}_{t_1},{#1}_{t_2},\ldots,{#1}_{t_n})}
\newcommand{\oph}[1]{(\op{#1}_{t_1},\op{#1}_{t_2},\ldots,\op{#1}_{t_n})}
\newcommand{\tp}[1]{\op{#1}_{t_1}\otimes\op{#1}_{t_2}\otimes\ldots\otimes\op{#1}_{t_n} }

\begin{document}
\begin{titlepage}
\hspace{8truecm} Imperial/TP/92-93/39

\begin{center}
                {\large\bf Quantum Logic and the Histories  \\[0.2cm]
                           Approach to Quantum Theory}
\end{center}
\vspace{1 truecm}
\begin{center}
        C.J.~Isham \\[0.5cm]
        Blackett Laboratory\\
        Imperial College\\
        South Kensington\\
        London SW7 2BZ\\
        United Kingdom
\end{center}

\begin{center} August 1993\end{center}

\begin{abstract}
An extended analysis is made of the Gell-Mann and Hartle axioms for a
generalised `histories' approach to quantum theory. Emphasis is placed
on finding equivalents of the lattice structure that is employed in
standard quantum logic. Particular attention is given to
`quasi-temporal' theories in which the notion of time-evolution is
less rigid than in conventional Hamiltonian physics; theories of this
type are expected to arise naturally in the context of quantum gravity
and quantum field theory in a curved space-time. The quasi-temporal
structure is coded in a partial semi-group of `temporal supports' that
underpins the lattice of history propositions. Non-trivial examples
include quantum field theory on a non globally-hyperbolic spacetime,
and a simple cobordism approach to a theory of quantum topology.

    It is shown how the set of history propositions in standard
quantum theory can be realised in such a way that
each history proposition is represented by a genuine projection
operator. This provides valuable insight into the possible lattice
structure in general history theories, and also provides a number
of potential models for theories of this type.

\end{abstract}
\end{titlepage}

\section{Introduction}
\subsection{Preliminary remarks}
A good illustration of the great difficulty encountered when
attempting to construct a fully comprehensive
account of quantum gravity is the striking lack of anything
remotely resembling an `axiomatic' scheme for such a theory. The point
at stake is not one of mathematical rigour {\em per se\/}, but
rather the need for a clear statement about what one is really trying
to do. This uncertainty is reflected in the wide range of declared views
on how to approach the subject and, indeed, on what
would constitute a successful resolution of the problem.

    The lack of any quasi-axiomatic scheme is regretable, but hardly
surprising, not least because of the frequently expressed view that
the well-developed space-time concepts of classical
general relativity may be inappropriate at scales of the Planck length
and time. One example is the well-known `problem of
time' that arises in attempts to construct a canonical theory of quantum
gravity
\footnote{For recent reviews see \cite{Kuc92a,Ish92}.}.
This problem first appears in the canonical approach to
classical general relativity where the notion of time as an external
parameter disappears and has to be recovered in some way from the
internal contents of the system. In the analogous quantum theory opinions
differ on whether (i) the concept of `time' is fundamental; or
(ii) our normal idea of time is a coarse-grained concept that
emerges only above the Planck scale in a theory of quantum gravity
that is fundamentally `timeless'. In the former case, a major issue is
whether the identification of an internal time should be made
before, or after, quantisation. Not unsurprisingly, the problem of
time is of major importance in quantum cosmology, particularly in
attempts to describe the origin of the universe as some type of
quantum-mechanical event.

    One of the more adventurous suggestions for resolving this issue
was made recently by Gell-Mann and Hartle in the context of the
`consistent histories' approach to quantum theory. The consistent
histories programme for standard quantum theory is currently
generating substantial interest, inspired mainly  by a major series of
papers by Gell-Mann and Hartle
\cite{GH90a,GH90b,GH90c,Har91a,Har91b,GH92,Har93a} which partly follow
on from, and are partly independent of, seminal work by Griffiths
\cite{Gri84} and by Omn\`es
\cite{Omn88a,Omn88b,Omn88c,Omn89,Omn90,Omn92}.

    This new perspective on quantum theory is of importance for a
variety of reasons, not least of which is the possibility it
affords of escaping the thralls of the
Copenhagen interpretation with its infamous measurement problem and
associated concept of state-vector reduction induced by an external
observer. Whether or not this hope is justified has been the subject of
much, sometimes acrimonious, debate, although most workers in the
field agree that, at the very least, the idea of consistent histories
provides valuable insight into the relation between classical physics
and quantum physics and, in particular, the sense in which the former
may be said to `emerge' from the latter.

    However, the motivation of the present paper is not a desire to
elude the clutches of Copenhagen epistemology but rather the
suggestion of Gell-Mann and Hartle to use the methodology of
consistent histories to tackle the problem of time in quantum gravity.
At a first glance, it seems odd to address the problem of time using
the notion of `history', a concept that is surely intrinsically
temporal? But Gell-Mann and Hartle refute this by asserting that a
`history' need not necessarily be regarded as a time-ordered string of
events or propositions: it could appear as a fundamental theoretical
entity in its own right.

    A simple example \cite{Har93a} of this idea is given by a
globally-hyperbolic Lorentzian metric $\g$ on a manifold $\M$. Any
such space-time can be expressed (in a variety of ways) as a
one-parameter $t$ (time) family of Riemannian geometries $g_t$ on a
spatial manifold $\Sigma$ with, topologically,
$\M\simeq\Sigma\times\R$. In this sense, $\g$ is a possible history of
the universe: \ie a path $t\mapsto g_t$ in the infinite-dimensional
space of 3-geometries on $\Sigma$. But suppose now that $\g$ is {\em
not\/} globally hyperbolic. The notion of time is still well-defined
locally on $(\cal M,\g)$ but, since there are no global time
functions, it is no longer possible to interpret $\g$ as a temporal
sequence of three-geometries; however, $\g$ may still be perfectly
acceptable as a space-time geometry, and hence as a `generalised
history' of the universe.  Note that a metric of this type
may still possess some spacelike hypersurfaces that divide $\M$ into
two pieces and, to that extent, some global time structure remains.
This is an example of what I shall call a {\em quasi-temporal\/}
situation.

    This example of Hartle is a mild one, but it illustrates the main
point well: in classical or quantum gravity a `history' is simply a
potential configuration for the universe as a whole, including any
quasi space-time structure it may possess.  This raises a number of
fascinating possibilities in which the class of possible universes is
deemed to be something considerably more exotic than a collection of
non globally-hyperbolic Lorentzian geometries. An example would be a
quantum topology theory in which the  histories include different
point-set topologies with the idea that the smooth manifold structure
of classical general relativity emerges only in some coarse-grained
sense.

\subsection{The main problems}
The main challenge, and the focus of the present paper, is to
translate these rather vague ideas into a concrete mathematical form.
What we seek is a quantum history framework that is as mathematically
well-formulated as are, for example, the various well-known
axiomatisations of standard quantum theory. Of course, any such
formalism must reproduce the standard theory where this is appropriate
but, hopefully, the axioms will also admit realisations that are very
different. This applies in particular to theories of quantum
space-time structure where the mathematical apparatus of standard
quantum theory (Hilbert spaces, self-adjoint operators, \etc) could
well be among the things that `emerge' at a suitably coarse-grained
level.

    The approach adopted in the present paper is motivated by the
presentation by Gell-Mann and Hartle of a set of generalised axioms
for a quantum theory based on histories and decoherence functionals.
My aim is to construct a refinement of these axioms using techniques
drawn from the `quantum logic' approach to standard quantum theory.
Thus we are looking for a history-based, quantum gravity analogue of
something like, for example, the Mackey \cite{Mac63} or Jauch-Piron
axioms \cite{Jau73,Pir76}. Following the seminal work of Von Neumann
and Birkhoff \cite{VNB36}, most approaches to quantum logic
\footnote{An excellent general introduction is \cite{BC81}.}
have concentrated on the lattice
\footnote{One can also employ the weaker concept of an
orthomodular partially ordered set; for a recent general exposition
see \cite{PP91}. It should be possible to develop a history analogue
of this using the ideas in the present paper.}
structure of the set $\L$ of propositions concerning the state of a
physical system at a {\em single\/} time. In realisations based on a
Hilbert space $\H$, the lattice $\L$ is identified with the set
$P(\H)$ of projection operators on $\H$ (or, equivalently, with the
set of all closed linear subspaces of $\H)$. Time evolution appears as
an automorphism of the lattice; in a Hilbert space model this is
implemented by the familiar family $t\mapsto U(t):=e^{-iHt}$ of
unitary operators. In classical physics the lattice of propositions is
a Boolean algebra, but in quantum theory this is no longer the case
since distributivity is lost: a feature that seems to accurately
reflect most of the peculiar features of the quantum world.

    The second ingredient in standard quantum logic is the space $\cal
R$ of quantum states. In a Hilbert space realisation this is
identified with the set of density matrices. The different
axiomatisations vary on whether the primary focus is placed on $\L$ or
on $\cal R$ but, in all cases, a dual relation is posited between
these spaces where the pairing $<\r,P>$ of a state $\r\in{\cal R}$
with a proposition $P\in\L$ is identified physically as the
probability that $P$ will be true if the state is $\r$.

    The proposal to use quantum logic in the consistent histories
formalism is not new; indeed, Omn\`es invoked such ideas in his
seminal work. However, his main aim was to show that any situation in
which histories decohere admits a sensible physical description using
standard Boolean logic, whereas my goal is quite different. I am not
concerned with the mechanism of decoherence as such---nor with the
interpretational nuances of quantum physics---but wish rather to
explore the basic mathematical structure that might be assigned to the
spaces of histories and decoherence functionals. Thus, in what
follows, references to quantum `logic' should {\em not\/} be read as
inferring that physical events must necessarily be described using a
non-standard form of logic; the implication is only that the set of
histories may possess a type of {\em algebraic\/} structure that is
similar to the lattices used in standard quantum logic. In particular,
I do {\em not\/} wish to become embroiled in the peristalithic debate
about whether logic itself is empirical or {\em a priori\/}!

    Most studies of quantum logic have involved propositions at a
single time and therefore are not directly applicable to the case in
hand. A notable exception is the work of Mittelstaedt and Stachow
\cite{Mit77,Mit78,Sta80,Sta81} who developed a general theory of the
logic of sequential propositions using sequential conjunctions of the
type `$A$ is true {\em and then\/} $B$ is true {\em and then\/}
$\ldots$', plus other temporal connectives such as `or then' and
`sequential implication'. One conclusion was that the
appropriate mathematical tool for describing sequential propositions
is a Baer-* semiring \cite{Sta80}: a mathematical structure that is
closely connected with what is done in standard quantum
theory using the history formalism.

    The approach adopted in the present paper can be considered in
some respects as an extension of the work of Mittelstaedt and Stachow
in a way that takes more account of the temporal labels and which
thereby admits an extension to quasi-temporal situations. A key
ingredient is the observation that the statement that a certain
universe (\ie history) is `realised'  is itself a proposition, and
therefore the set of all such histories might possess a lattice
structure analogous to the lattice of single-time propositions
in standard quantum logic. In particular, a history
proposition might be representable by a {\em projection\/} operator in
some Hilbert space.

    It must be emphasised that representing histories with projection
operators is precisely what does {\em not\/} happen in the usual
approach to standard quantum theory. The central object of interest is
the so-called {\em decoherence functional\/} (see \eq{Def:d_rho}), in
whose calculation the history $\a$ corresponding to single-time
propositions $\a_{t_1},\a_{t_2},\ldots,\a_{t_n}$ at times
$t_1<t_2<\ldots<t_n$ is associated with the `$C$-representation'
defined as $\op{C}_\a:=$
$\op\a_{t_n}(t_n)\op\a_{t_{n-1}}(t_{n-1})\ldots\op\a_{t_1}(t_1)$: a
product of (Heisenberg picture) projection operators
$\op\a_{t_k}(t_k)$ that is usually not itself a projection operator.
This confuses the logical analysis of the system. For example, if $\a$
and $\b$ are disjoint
\footnote{This notion will be defined properly in Section
\ref{Sec:Search}}
histories the $C$-representative $\op{C}_{\a\vee\b}$  of `history $\a$
{\em or\/} history $\b$' is always taken to be $\op{C}_\a+\op{C}_\b$:
a claim that is usually justified by an appeal to a path-integral
formalism. However, a deeper analysis might want to deal with the
problem in a more realisation-independent way, in which case the
statement $\op{C}_{\a\vee\b}=\op{C}_\a+\op{C}_\b$ is reminiscent
of the standard quantum-logic result that if $P$ and $Q$ are
single-time, disjoint propositions, the projection operator that
represents `$P$ {\em or\/} $Q$' is $\P+\Q$. But the assignment of
$\op{C}_\a+\op{C}_\b$ cannot be justified in this simple way if
$\op{C}_\a$ or $\op{C}_\b$ are not projection operators.

    This problem becomes more acute if one asks for the
$C$-representation of `$\a$ {\em or\/} $\b$' when $\a$ and $\b$ are
{\em not\/} disjoint. The analogous question in quantum logic has a
direct answer---the proposition $P\vee Q$ is represented by the
operator that projects onto the closure of the linear span of the
subspaces onto which $\P$ and $\Q$ project---but the answer for
histories is not obvious. A similar query attends the common
identification of $\op1-\op{C}_\a$ as the $C$-representative of the
proposition that the history is  `{\em not\/} $\a$'

    The challenge in construcing a full axiomatic scheme for a quantum
history structure is four-fold:
\be
\item Find a general set of axioms for a quantum history theory that
has the same degree of precision as one of the well-known axiomatic
schemes for standard Hamiltonian quantum theory. This scheme should
be broad enough to admit quantum gravity realisations in which the
classical notions of space and time are not applicable.

\item If, as expected, the set of all history propositions has some
sort of lattice structure, study the way in which this differs from
the standard quantum-logic lattices of fixed-time propositions. In
particular, what role is played by any `quasi-temporal' properties of
the system?

\item Study the extent to which history propositions can be
represented as projection operators on some Hilbert space.

\item Study general ways of constructing realisations of the history
axioms.
\ee

    One of the tasks of the present paper is to construct a history
version of standard quantum theory in which every history {\em is\/}
represented by a projection operator
\footnote{To avoid confusion it must be emphasised that the Hilbert
space on which the histories are represented by projection operators
is {\em not\/} the normal Hilbert space of the quantum theory.}.
The usual logical operations on projection operators can then be
applied to place a lattice structure on the set of all
history propositions. This provides some justification for asserting
that, in any history theory, the set of propositions about the
histories should have the structure of a non-distributive lattice, of
which good models may be the lattices of projection operators in a
Hilbert space.

    Of course, the challenge of quantum gravity may require
realisations of the general history axioms exist that do not involve
Hilbert spaces in any way at all. On the other hand,
powerful reconstruction theorems in standard quantum theory show
that if the lattice $\L$ of single-time propositions satisfies certain
plausible conditions it must necessarily be realised as the lattice of
subspaces of some (not necessarily complex) Hilbert space. To avoid
the analogue of this result in the history case, the algebraic
properties of the set of history propositions and the associated space
of decoherence functionals would have to differ in some significant
way from those of the standard lattice of single-time propositions and
associated space of states.

    The plan of the paper is as follows. We start with a short
exposition of the way histories are used in standard Hamiltonian
quantum theory, and of the axioms proposed by Gell-Mann and Hartle.
Then the search for a general set of history axioms begins with the
introduction of the notion of a {\em temporal support\/} for use in a
theory that may possess only quasi-temporal properties. This is based
on the observation that one of the most prominent features of a
certain type of history (a {\em history filter\/}: a sort of
sequential conjunction) in standard quantum theory is that it can be
decomposed  into sub-histories in which one sub-history `follows'
another. The claim is that the associated structure of a partial
semi-group is still present in the general case where it encodes
whatever quasi-temporal properties may be possessed by the system.

    It is shown that the space of history filters possesses a natural
lattice structure, but this is not the correct one physically and
only the $\wedge$ semi-lattice sub-structure can be retained. It is
necessary therefore to embed the space of history filters into a
larger space of general `history propositions' that carries the correct
lattice structure. To validate this crucial point, section
\ref{Sec:HPO} is devoted to the construction of a realisation of the
logical structure of standard quantum theory in which {\em all\/}
history propositions are represented by projection operators on a new
Hilbert space. These results are then used to motivate a tentative set
of axioms for a general history theory with temporal supports.
Relativistic quantum field theory is discussed briefly in section
\ref{Sec:QFT}, mainly as a way of illustrating the use of temporal
supports in a quasi-temporal theory. The paper concludes with a short
account of some of the outstanding problems for further research.

\section{The Search for General History Axioms}
\label{Sec:Search}
\subsection{Histories in standard quantum theory}
It will be helpful to begin by summarising
very briefly how `histories' are understood in the conventional
interpretation of an open, Hamiltonian quantum system that is subject
to measurements by an external, and essentially classical, observer.

    To this end, let $U(t_1,t_0)$ denote
\footnote{For typographical ease, `hats' will not be placed on
operators until it becomes necessary to distinguish propositions from
the projection operators that represent them.}
the unitary time-evolution operator from time $t_0$ to $t_1$; \ie
$U(t_1,t_0)=e^{-i(t_1-t_0)H/\hbar}$. Then, in the Schr\"odinger
picture, the density operator state $\r(t_0)$ at time $t_0$ evolves in
time $t_1-t_0$ to $\r(t_1)$, where
 \beq
\r(t_1)=U(t_1,t_0)\r(t_0)U(t_1,t_0)^{\dag}=U(t_1,t_0)\r(t_0)U(t_1,t_0)^{-1}.
\eeq
Suppose that a measurement is made at time $t_1$ of a property
represented by a projection operator $P$. Then the probability that
the property will be found is
\beq
{\rm Prob}\big(P=1;\rho(t_1)\big)=\tr\big(P\r(t_1)\big) =
    \tr\big(PU(t_1,t_0)\r(t_0)U(t_1,t_0)^{\dag}\big) =
        \tr\big(P(t_1)\r(t_0)\big)
\eeq
where
\beq
    P(t_1):=U(t_1,t_0)^{\dag}P(t_0)U(t_1,t_0)
\eeq
is the Heisenberg picture operator defined with respect to the fiducial
time $t_0$. If the result of this measurement is kept then, according
to the Von Neumann-L\"uders `reduction' postulate, the appropriate
density matrix to use for any further calculations is
\beq
\r_{\rm red}(t_1):={P(t_1)\r(t_0)P(t_1)\over\tr\big(P(t_1)\r(t_0)\big)}.
\eeq

    Now suppose a measurement is performed of a second observable $Q$
at time $t_2>t_1$. Then, according to the above, the {\em
conditional\/} probability of getting $Q=1$ at time $t_2$ given that
$P=1$ was found at time $t_1$ (and that the original state was
$\r(t_0)$) is
\beq
    {\rm Prob}\big(Q=1|P=1{\rm\ at\ }t_1;\r(t_0)\big)=
        \tr\big(Q(t_2)\r_{\rm red}(t_1)\big)=
    {\tr\big(Q(t_2)P(t_1)\r(t_0)P(t_1)\big)\over\tr\big(P(t_1)\r(t_0)\big)}.
\eeq
The probability of getting $P=1$ at $t_1$ {\em and\/} $Q=1$
at $t_2$ is this conditional probability multiplied by ${\rm
Prob}\big(P=1;\r(t_1)\big)$, \ie
\beq
{\rm Prob}\big(P=1{\rm\ at\ }t_1{\rm {\em\ and\ }}Q=1{\rm\ at\ }t_2;
    \r(t_0)\big)=\tr\big(Q(t_2)P(t_1)\r(t_0)P(t_1)\big).
\eeq
Generalising to a sequence of measurements of propositions
$\a_{t_1},\a_{t_2},\ldots,\a_{t_n}$ at times $t_1,t_2,\ldots,t_n$, the
joint probability of finding all the associated properties is
\beqa
\lefteqn{{\rm Prob}\big(\a_{t_1}=1{\rm\ at\ }t_1 {\rm{\em\ and}\ }
    \a_{t_2}=1{\rm\ at\ }t_2{\rm{\em\ and\ldots}}
    \a_{t_n}=1{\rm\ at\ }t_n;\r(t_0)\big)=\ \ \ \ }           \nonumber\\
	&&\tr\big(\a_{t_n}(t_n)\ldots\a_{t_1}(t_1)\r(t_0)
		\a_{t_1}(t_1)\ldots\a_{t_n}(t_n)\big)
                                                    \label{Prob:a1_an}
\eeqa
where we have used the relation $P^2=P$ for a projection operator.

    The main assumption of the consistent histories interpretation of
quantum theory is that, under appropriate conditions, the probability
assignment \eq{Prob:a1_an} is still meaningful for a {\em closed\/}
system, with no external observers or associated measurement-induced
state-vector reductions (thus signalling a move from `observables' to
`beables'). The satisfaction or otherwise of these conditions is
determined by the behaviour of the {\em decoherence functional\/}
$d_\r(\a,\b)$ which, for the pair of sequences of projection operators
$\a:=\h{\a}$  and $\b:=\h{\b}$ is defined as
\beq
  d_\r(\a,\b):=\tr\big(\a_{t_n}(t_n)\ldots\a_{t_2}(t_2)\a_{t_1}(t_1)\r(t_0)
                     \b_{t_1}(t_1)\b_{t_2}(t_2)\ldots\b_{t_n}(t_n)\big)
        =\tr\big(C_\a\rho(t_0)C_\b^{\dag}\big)  \label{Def:d_rho}
\eeq
where
\beqa
    C_\a&:=&\a_{t_n}(t_n)\ldots\a_{t_2}(t_2)\a_{t_1}(t_1)\nonumber\\
        &=&U(t_0,t_n)\a_{t_n}U(t_n,t_{n-1})\ldots
          U(t_3,t_2)\a_{t_2}U(t_2,t_1)\a_{t_1}U(t_1,t_0)
                                                \label{Def:C_a}
\eeqa
The conditions required for the probability assignments
\eq{Prob:a1_an} to be consistent are summarised below in the (more
general) context of the Gell-Mann and Hartle axioms.

    It is useful at this point to gather together a number of
definitions that are associated with these ideas.
\bi
\item  A {\em homogeneous history\/}
\footnote{The justification of the term `homogeneous' will be given
in section \ref{SSec:HPO_fixedTS}.})
is any time-ordered sequence $\oph{\a}$ of
projection operators.

\item  A homogeneous history  $\b:=\oph{\b}$ is {\em coarser\/} than
another history $\a:=\oph{\a}$ if, for every $t_i$,
$\op\b_{t_i}<\op\a_{t_i}$ where $<$ denotes the usual ordering
operation on the space of projection operators, \ie $\P<\Q$ means that
the range of $\P$ is a subspace of the range of $\Q$ (this includes
the possibility that $\P=\Q$ so that, in particular, every homogeneous
history is trivially coarser than itself). This relation on the set of
homogeneous histories is a partial ordering
 \footnote{A relation $<$ on a set $X$ is a {\em partial ordering\/}
if it satisfies the conditions (i) for all $x\in X$, $x<x$; (ii) $x<y$
and $y<x$ implies $x=y$; and (iii) $x<y$ and $y<z$ implies $x<z$.}.

\item Two homogeneous histories  $\a:=\oph{\a}$ and $\b:=\oph{\b}$ are
{\em disjoint\/} if, for at least one time point $t_i$, $\op\b_{t_i}$
is disjoint from $\op\a_{t_i}$, \ie the ranges of these two projection
operators are orthogonal subspaces of $\H$.

\item The {\em unit\/} history assigns the unit operator to any time
$t$; the {\em null\/} history assigns the zero operator.

\item In calculating a decoherence functional it may be necessary to
go outside the class of homogeneous histories to include {\em
inhomogeneous\/} histories. A history of this type arises as a
logical `or' (denoted $\vee$) operation on a pair of disjoint
homogeneous histories  $\a:=\oph{\a}$ and $\b:=\oph{\b}$. Such a
history $\a\vee\b$ is generally {\em not\/} itself a collection of
projection operators
\footnote{The exception is when $\a$ and $\b$ differ only in their
values $\a_{t_i}$ and $\b_{t_i}$ at a single time point $t_i$. In this
case, if the single-time projection operators $\a_{t_i}$ and
$\b_{t_i}$ are disjoint then $\a\vee\b$ is equal to $\a$ at all time
points except $t_i$, where it is $\a_{t_i}+\b_{t_i}$. In particular,
$\a\vee\b$ is still homogeneous.}
(\ie it is not homogeneous) but, when computing the decoherence
functional, it is represented by the operator
$\op{C}_{\a\vee\b}:=\op{C}_\a+\op{C}_\b$. The coarse-graining
relations $\a<\a\vee\b$ and $\b<\a\vee\b$ are deemed to apply to this
disjoint `or' operation. The `negation' operation $\neg$ also usually
turns a homogeneous history into an inhomogeneous history, with
$\op{C}_{\neg\a}:=\op{1}-\op{C}_\a$.
\ei

    Note that the collection $\oph{\a}$ of projection operators can
itself be viewed as a projection operator on the  direct sum
$\oplus_{t\in\{t_1,\ldots,t_n\}}\H_t$ of $n$ copies of the Hilbert
space $\H$, with the copies being labelled by the time values
$\{t_1,t_2,\ldots,t_n\}$ at which $\a$ is defined. This idea will be
generalised in section \ref{Sec:HPO} with a construction in which
{\em every\/} history, inhomogeneous as well as homogeneous, is
represented by a projection operator on a certain Hilbert space (albeit,
not the direct sum $\oplus_{t\in\{t_1,\ldots,t_n\}}\H_t$).

\subsection{The Gell-Mann and Hartle axioms}
\label{SSec:HarAxioms}
The Gell-Mann and Hartle axioms \cite{Har93a} generalise the
discussion above by postulating a new approach to quantum theory in
which the notion of `history' is ascribed a fundamental role; \ie a
history may be an irreducible entity in its own right that is not
necessarily to be construed as a time-ordered string of single-time
propositions. For our purposes, these axioms and definitions are
essentially as follows:
\be
\item The fundamental ingredients in the theory are a space of {\em
histories\/} and a space of {\em decoherence functionals\/} which are
complex-valued functions of pairs of histories.

\item The set of histories possesses a partial order $<\,$. If
$\a<\b$ then $\b$ is said to be {\em coarser\/} than $\a$, or a {\em
coarse-graining\/} of $\a$; dually, $\a$ is a {\em finer\/} than $\b$,
or a {\em fine-graining\/} of $\b$ . Heuristically this means that
$\a$ provides a more precise specification than $\b$.

\item There is a notion of two histories $\a,\b$ being {\em
disjoint\/}, written $\a\perp\b$. Heuristically, if
$\a\perp\b$ then if either $\a$ or $\b$  is  `realised', the other
is automatically excluded.

\item There is a {\em unit\/} history $1$ (heuristically, the history
that is always realised) and a {\em null history\/} $0$
(heuristically, the history that is never realised). For all histories $\a$
we have $0<\a<1$.

\item Two histories $\a,\b$ that are disjoint can be
combined to form a new history $\a\vee\b$ (heuristically, the history
`$\a$ {\em or\/} $\b$').

\item A set of  histories $\a^1,\a^2,\ldots,\a^N$ is said to be
{\em exclusive\/} if $\a^i\perp\a^j$ for all $i,j=1,2,\ldots,N$. The set
is {\em exhaustive\/} (or {\em complete\/}) if it is exclusive and if
$\a^1\vee\a^2\vee\ldots\vee\a^N=1$.

\item Any decoherence functional $d$ satisfies the following
conditions:
    \be
    \item $d(0,\a)=0$ for all $\a$.
    \item {\em Hermiticity\/}: $d(\a,\b)=d(\b,\a)^*$ for all
          $\a,\b$.
    \item {\em Positivity\/}: $d(\a,\a)\ge0$ for all $\a$.
    \item {\em Additivity\/}: if $\a\perp\b$ then, for
          all $\g$, $d(\a\vee\b,\g)=d(\a,\g)+d(\b,\g)$.
    \item {\em Normalisation\/}: If $\a^1,\a^2,\ldots,\a^N$ and
          $\b^1,\b^2,\ldots,\a^M$ are two complete sets of
          histories then
          \beq
                \sum_{i=1}^N\sum_{j=1}^Md(\a^i,\b^j)=1.
          \eeq
    \ee
\ee
\noindent
In standard quantum theory, the definition \eq{Def:d_rho}
automatically satisfies all these conditions
although the discussions in the literature are sometimes a
little vague about the  extent to which the conditions apply
to histories that are inhomogeneous in addition to those that are
homogeneous.

    It is important to note that this axiomatic scheme is given a
physical interpretation only in relation to {\em consistent\/} sets of
histories. A complete set $\cal C$ of histories is said to be
(strongly) consistent with respect to a particular decoherence
functional $d$ if $d(\a,\b)=0$ for all $\a,\b\in{\cal C}$ such that
$\a\ne\b$. Under these circumstances, $d(\a,\a)$ is given the physical
interpretation as the {\em probability\/} that the history $\a$ will
be `realised'
\footnote{Deciding on the precise meaning of `realised' is part of the
general ontological problem that I wish to avoid discussing in this
particular paper.}.
The Gell-Mann and Hartle axioms then guarantee that the usual
Kolmogoroff probability sum rules will be satisfied.

    The only major difference between the axioms above and those
actually given by Gell-Mann and Hartle is that the latter (i) extend
the above to include {\em infinite\/} sums; and (ii) invoke a notion
of `fine-grained' histories. The former may well be desirable but it
first requires defining what is meant by taking a limit of histories;
we shall return briefly to this issue later. A history $\a$ is defined
to be {\em fine-grained\/} if the only histories $\b$ for which
$\b<\a$ are $0$ or $\a$ itself. In standard quantum theory, histories
of this type arise as time-ordered sequences of projection operators
whose ranges are all one-dimensional subspaces of the Hilbert space.
However, it is not clear whether the existence of such entities should
postulated in general
\footnote{Even in a Hilbert space situation, `atoms' of this type would
not be present if we were concentrating on histories associated with
observables whose spectra was purely continuous.}  or, if so, what
additional properties they should possess. A particularly advantageous
situation is when the space of histories is {\em atomic\/} in the
sense that there exist so many fine-grained, and disjoint, histories
that each history can be written as a unique $\vee$-sum of these
atoms. In this case, the additivity property means that the value of a
decoherence functional is determined uniquely by its values on pairs
of atoms.

\subsection{History filters and temporal supports in standard quantum
logic}
\label{SSec:HisFil}
We now begin our detailed development of the Gell-Mann and Hartle
axioms with the aim of providing a general quantum-logic theory of
histories that incorporates whatever quasi-temporal properties the
system may possess. In what follows, the construction of such a
general scheme will be motivated by the special example of a history
version of standard quantum logic. By this is meant a generalisation
of the ideas above in which strings of projection operators are
replaced by strings of single-time propositions belonging to the
lattice $\L$ of some `standard' quantum logic theory. The hope is that
the algebraic structure thus revealed will provide valuable clues to
what should be postulated in the general case.

    In standard quantum logic the lattice $\L$ of single-time
propositions is highly structured. In particular, it is usually
required to be {\em complete\/},
\footnote{A lattice is {\em complete\/} if the meet and join
operations can be applied to arbitrary subsets of elements. It is {\em
countably complete\/} if this can be done for countable subsets.}
{\em orthocomplemented\/}
\footnote{A lattice is {\em orthocomplemented\/} if there is a map
$\a\mapsto\neg\a$ with the property that (i) $\neg(\neg\a)=\a$, (ii)
$\a<\b$ implies $\neg\b<\neg\a$, and (iii) $\a\wedge\neg\a=0$ and
$\a\vee\neg\a=1$.},
{\em orthomodular\/}
\footnote{A lattice is {\em orthomodular\/} if $\a<\b$ implies that
$\b=\a\vee(\b\wedge\neg\a)$.},
{\em atomic\/}
\footnote{An element $\a$ of a lattice is an {\em atom\/} if
$0<\b<\a$ implies that $\b=0$ or $\b=\a$. A lattice is {\em atomic\/}
if for any non-zero element $\b$ there exists an atom $\a$ with
$\a<\b$.}
and with the {\em covering\/} property
\footnote{An element $\g$ {\em covers\/} $\b$ if $\b<\g$ and
$\b<\d<\g$ implies that $\d=\b$ or $\d=\g$. An atomic lattice has
the {\em covering property\/} if for every $\b$ and every atom $\a$
such that $\a\wedge\b=0$ the element $\a\vee\b$ covers $\b$.}.
A {\em state\/} on the lattice $\L$ is defined to be a real-valued
function $\mu$ on $\L$ that is a `probability measure' \ie
\beq
    0\le\mu(\a)\le1 {\rm\ for\ all\ }\a\in\L,  \label{Def:state_1}
\eeq
\beq
    \mu(0)=0 {\rm\ and\ }\mu(1)=1            \label{Def:state_2}
\eeq
and the countable additivity property that if $\a^1,\a^2,\ldots$ is a
finite or countable sequence of elements in $\L$ that are pairwise
disjoint
\footnote{Two propositions $\a$ and $\b$ are {\em disjoint\/} if $\a<\neg\b$.}
then the series $\mu(\a^1\vee\a^2\vee\ldots)$ converges and
\beq
    \mu(\a^1\vee\a^2\vee\ldots)=\mu(\a^1)+\mu(\a^2)+\ldots.
                                                \label{Def:state_3}.
\eeq
Clearly the space $\cal R$ of all states is a real convex set, \ie if
$\mu$ and $\nu$ are states then so is $r\mu+(1-r)\nu$ for all real $r$
with $0<r<1$.

    One of the main aims of this paper is to find candidates for the
`history analogues' of the lattice $\L$ and the state-space $\cal R$
of the standard Hamiltonian theory. These analogues of $\L$ and
$\cal R$ will be a space $\UP$ of history-propositions and a space
$\cal D$ of decoherence functionals.

    We begin with a careful account of some aspects of the
quantum-logic version of the standard history theory sketched above.
Thus we consider a system with a lattice $\L$ of single-time
propositions and start by defining a {\em history filter\/} to be any
finite
\footnote{Circumstances might arise where it is necessary to consider
{\em infinite\/} collections of propositions. This is awkward within a
purely algebraic formalism but one way of handling this
problem will appear later in section \ref{Sec:HPO}.}
collection $\h{\a}$ of single-time propositions $\a_{t_i}\in\L$ which is
time-ordered in the sense that $t_1<t_2<\ldots<t_n$. Thus, in the
special case where $\L$ is identified with the lattice $P(\H)$ of
projection operators on a Hilbert space $\H$, a history filter
is what we earlier called a homogeneous history. However, to
distinguish the general situation from the special
case, I will continue to refer to a time-ordered string of
projection operators as a `homogeneous history' and  reserve the
phrase `history filter' for either the quantum-logic version of standard
quantum theory, or for the general quantum-logic theory.

    In the case of standard quantum logic, a history filter is a
time-labelled version of what Mittelstaedt and Stachow call a {\em
sequential conjunction\/}\cite{Mit77,Sta80,Sta81} \ie it corresponds
to the proposition `$\a_{t_1}$ is true at time $t_1$, {\em and then\/}
$\a_{t_2}$ is true at time $t_2$, {\em and then\/} $\ldots$ {\em and
then\/} $\a_{t_n}$ is true at time $t_n$'. The phrase `history filter'
itself is intended to capture the idea that each single-time
proposition $\a_{t_i}$ in the collection $\h{\a}$ serves to `filter
out' the properties of the system that are realised in the history of
the universe.

    It is important to be able to manipulate history filters that are
associated with {\em different\/} sets of time points. To this end, it
is useful to think of a history filter as something that is defined at
{\em every\/} time point but which is `active' only at a finite subset
of points. This can be realised mathematically by defining it to be
equal to the trivial proposition at all but the active points. More
precisely, in standard quantum logic we shall define a history filter
$\a$ to be an element of the space ${\cal F}(\T,\L)$ of maps from the
space of time points $\T$ (in the present case, the real line $\R$) to
the lattice $\L$ with the property that each map is (i) equal nowhere
to the null single-time proposition, and (ii) equal to the unit
single-time proposition for all but a finite set of $t$ values. It
will be convenient to append to this space the null history filter
which is defined to be the null single-time proposition at all points
$t\in\T$.
\footnote{It is also sometimes convenient to allow history filters
that are equal to the null single-time proposition at one or more time
points but are non-null at other points: these are all regarded as
equivalent to the null history.}

    It follows that, in a standard quantum theory realised on a
Hilbert space $\H$, a history filter (\ie a homogeneous history) is
represented by an element $\a$ of the space of functions ${\cal
F}(\T,P(\H))$ where $\op\a_t$ (the value of the map $\a$ at $t\in\T$)
is equal to the unit operator for all but a finite set of time points
$t\in\T$. Objects of this type can be regarded as projection operators
on the weak direct sum ${\cal F}(\T,\H)$ of $\H$-valued functions on
$\T$.
\footnote{The qualification `weak' refers to the fact that the
functions $v:\T\map\H$ in ${\cal F}(\T,\H)$ are equal to the null
vector in $\H$ for all but a finite set of $t$ values. The scalar
product between two such vectors $v,w$ is defined by
$<v,w>:=\sum_{t\in\T}<v_t,w_t>_{\H}$ where $<,>_{\cal H}$ denotes the
scalar product on the Hilbert space $\H$.}

    The first of my suggested axioms for a general history theory is
that it should be based on a space of history filters, but these will
no longer just be time-ordered sequences of single-time propositions.
In the context of quantum cosmology, a history filter is a possible
`{\em universe\/}' complete with whatever quasi-temporal attributes it
may, or may not, possess. For this reason, the set of all history
filters in the general theory will be denoted $\U$; in the case of
standard quantum logic we will write $\U(\L):={\cal F}(\T,\L)$ to
indicate the underlying lattice $\L$ of single-time propositions.

    Any quasi-temporal properties of the system should be coded in the
mathematical structure of $\U$. Thus we must decide how the
propositional structure of the set of history filters can be
distinguished from that of a set of fixed-time propositions.

    The example of standard Hamiltonian quantum theory suggests that
the most characteristic property of a history filter is the
possibility of dividing it into two sub-histories, one of which
`follows' the other. Conversely, if a history filter $\b$ follows
another one $\a$, they can be combined to give a new filter
$\a\circ\b$ that, {\em pace\/} Mittelstaedt and Stachow, can be read
as `history $\a$ {\em and then\/} history $\b$'. More precisely, we
say that a history filter $\b:=(\b_{t_1'},\b_{t_2'}\ldots\b_{t_m'})$ {\em
follows\/} $\a:=(\a_{t_1},\a_{t_2}\ldots\a_{t_n})$ if $t_n<t_1'$, and
then define the combined history $\a\circ\b$ as
\footnote{A slight variant is to say that $\b$ follows $\a$ if
(i) $t_n<t'_1$, or (ii) $t_n=t_1'$, in which case we must have
$\a_{t_n}=\b_{t'_1}$. Then define $\a\circ\b$ by
\eq{Def:a_circ_b} if $t_n<t'_1$, and by
$\a\circ\b:=(\a_{t_1},\a_{t_2},\ldots,\a_{t_n},\b_{t'_2},\ldots,\b_{t'_m})$
if $t_n=t'_1$.}
\beq
    \a\circ\b:=(\a_{t_1},\a_{t_2},\ldots,\a_{t_n},
              \b_{t'_1},\b_{t'_2},\ldots,\b_{t'_m}).\label{Def:a_circ_b}
\eeq
It is clear that this operation satisfies the associative law
\beq
    \a\circ(\b\circ\g)=(\a\circ\b)\circ\g       \label{a(bg)=(ab)g}
\eeq
whenever the history filters $\a$, $\b$ and $\g$ are so related that
both sides of this equation are well-defined. Thus the space $\U(\L)$ of
history filters in standard quantum logic becomes a {\em partial
semi-group\/}
\footnote{A set $X$ equipped with a combination law $\circ$ is a {\em
semi-group\/} if the combination is associative, \ie $x\circ(y\circ
x)=(x\circ y)\circ z$ for all $x,y,z\in X$. It is a {\em partial\/}
semi-group if the combination law can be applied only to certain pairs
of elements.}
with respect to the combination law $\circ$.

    It is clear that the temporal properties of a history filter
$\a\in\U(\L)$ are encoded in the finite set of time points at which it
is active; \ie the points $t\in\T$ such that $\a_t\ne 1$. This
motivates the following definitions:
\be

\item The set of $t\in\T$ for which $\a_t\ne1$ is called the {\em
      temporal support\/}, or just {\em support\/}, of $\a\in\U(\L)$,
      and is denoted $\s(\a)$.

\item The set of all possible temporal supports will be denoted $\S$;
      in the present case this is just the set of all finite subsets
      of $\T=\R$.

\item The support of the null history is defined to be the empty
      subset of $\R$.
\ee

    Whether or not a history filter $\b$ follows another history
filter $\a$ is determined solely by their temporal supports. In fact,
the space $\S$ of supports of standard Hamiltonian theory can itself be
equipped with the structure of a partial semigroup by saying that the
support $s_2:=\{t'_1,t'_2,\ldots,t'_m\}$  {\em follows\/} the support
$s_1:=\{t_1,t_2,\ldots,t_n\}$ if $t_n<t'_1$, and then defining the
composition as
\beq
    s_1\circ s_2:=\{t_1,t_2,\ldots,t_n,t'_1,t'_2,\ldots,t'_m\}.
\eeq

    The relation between the partial semi-group structures on $\U(\L)$ and
$\S$ is captured by the observation that the support map
$\s:\U(\L)\map{\cal S}$ is a {\em homomorphism\/} between these two
partial semi-groups, \ie
\beq
    \s(\a\circ\b)=\s(\a)\circ\s(\b)         \label{s(a_circ_b)=}
\eeq
whenever both sides of the equation are defined.

    As a matter of convention we shall define the null support to
follow, and preceed, {\em every\/} element $s\in\S$ so that
$\a\circ0$, $0\circ\a$, $\a\circ1$ and $1\circ\a$ are defined for all
$\a\in\U(\L)$ with values $\a\circ0=0=0\circ\a$ and
$\a\circ1=\a=1\circ\a$ respectively. Thus the unit history $1$ serves
as a unit for the semi-group structure of $\U(\L)$ while the null
history $0$ is an absorbing element.

\subsection{History filters and temporal supports in the general case}
\label{SSec:SigTS}
The point of the discussion above is to emphasise that the essential
temporal properties of the space of history filters $\U(\L)$ in standard
Hamiltonian quantum theory is reflected in its partial semi-group of
supports. This raises the possibility of constructing a
`quasi-temporal' theory that is more general than the standard theory
and in which the quasi-temporal stucture is reflected in the structure of
the support space; a theory that was totally `non-temporal' would have
a support space equal to a single point.

    Thus we postulate that, in the general case,  the space $\U$ of
history filters (\ie possible universes) has the structure of a {\em
partial semi-group\/}. If $\a$ and $\b$ are such that they can be
combined to form $\a\circ\b$ we shall say that $\a$ {\em preceeds\/}
$\b$ (written as $\a\lhd\b$), or $\b$ {\em follows\/} $\a$. Broadly
speaking, the idea is that if $\b$ follows $\a$ then there is some
causal influence (perhaps, in a rather unusual sense) of  events
`localised' in the history $\a$ on those localised in $\b$. This
concept of `following' is captured in a  space $\S$ of {\em temporal
supports\/} that is associated with $\U$ and which is also a partial
semigroup; the two are related by a homomorphism $\s:\U\map\S$ as in
\eq{s(a_circ_b)=}.

    Note that a homomorphism is generally a many-to-one map, and
therefore information about a history filter $\a$ is lost when it is
mapped to $\s(\a)\in\S$. The key idea is that its image
in $\S$ just captures the essential information about any
quasi-temporal properties it may possess. This raises the
prospect of constructing a {\em tower\/} of semi-groups that can interpolate
between $\U$ and $\S$. Such a tower would be a series of partial semigroups
$\U_1,\U_2,\ldots,\U_k$ and homomorphisms $\s_i:\U_i\map\U_{i+1}$,
with $\s_0:\U\map\U_1$ and $\s_k:\U_k\map\S$, such that the support
homomorphism $\s:\U\map\S$ factorises into the finite chain
\beq
        \U{\buildrel\s_0\over\rightarrow}\U_1
          {\buildrel\s_1\over\rightarrow}\U_2
          {\buildrel\s_2\over\rightarrow}\ldots
          {\buildrel\s_{k-1}\over\rightarrow}\U_k
          {\buildrel\s_k\over\rightarrow}\S.    \label{tower}
\eeq
each element of which contains less information than the previous one
in the chain.

    It should be emphasised that the relation of `following' may not
be transitive; \ie $\a\lhd\b$ and $\b\lhd\g$ need not imply that
$\a\lhd\g$; in particular, we do not suppose that $\lhd$ is a partial
order on $\U$. The reason is that the set of events `localised' in
$\b$ that are causally affected by events in $\a$ need not include any
events that causally affect events in $\g$; an explicit example of
this phenomenon will appear in the discussion in section \ref{Sec:QFT}
of quantum field theory in a non globally-hyperbolic, curved
space-time where the support structure is different from that of
standard Hamiltonian theory.  We should also not rule out the
possibility that a history filter may precede itself, or the existence
of pairs of histories $\a,\b$ such that $\a\lhd\b$ and $\b\lhd\a$,
leading to a sort of `closed time-like loop'.

    A more exotic example of a potential support structure is given by
the partial semi-group of cobordism equivalence classes. More
precisely, given three closed $n$-manifolds $\Sigma_1$, $\Sigma_2$ and
$\Sigma_3$, equivalence classes $c_1$ and $c_2$ of cobordisms
\footnote{A {\em cobordism\/} from $\Sigma_1$ to $\Sigma_2$ is an
$n+1$-dimensional manifold $\M$ whose boundary is the disjoint union
of submanifolds diffeomorphic to $\Sigma_1$ and $\Sigma_2$.}
from $\Sigma_1$ to $\Sigma_2$ and $\Sigma_2$ to $\Sigma_3$
respectively can be combined to give a well-defined \cite{Mil65}
equivalence class $c_1\circ c_2$ of cobordisms from $\Sigma_1$ to
$\Sigma_3$. If desired, a `time-direction' can be introduced by
declaring one of the pair of boundaries $\Sigma_1,\Sigma_2$ in a
cobordism triple $(\Sigma_1,\Sigma_2,\M)$ to be the `initial'
$n$-space and the other to be the `final' $n$ space, with the
requirement that the combination $c_1\circ c_2$ of cobordisms is
permitted only if the final $n$-space of $c_1$ is diffeomorphic to the
initial $n$-space of $c_2$; \ie this is the condition that $c_2$
follows $c_1$. A temporal-support structure based on this cobordism
operation would be a natural foundation for a certain type of `quantum
topology' theory. This is another example where the relation
of `following' is clearly not transitive. This is shown in Figure 1 where
$B$ follows $A$ and $C$ follows $B$ but $C$ does not follow (\ie is
not cobordant with) $A$ (note that the `final' boundary of $C$ is the
empty set.)

    It is worth noting in passing that cobordism theory produces a
{\em category\/} in which the objects are closed manifolds and the
morphisms are equivalence classes of cobordisms. In fact, category
theory is the most general language within which to discuss temporal
supports. More precisely, the generic model for a space of temporal
supports is based on the morphisms in a category that is small (\ie
the collection of objects is a set rather than a class); this is also
related to the general theory of groupoids \cite{Hig71,Mac87}. This
idea (in particular, the example of cobordism theory) will be
developed in a later paper.

    In summary, the discussion above suggests that a general
history theory will contain the following ingredients.
\be
\item The basic entity in the theory is a partial semi-group $\U$ of
      history filters (\ie possible `universes').

\item There is an associated partial semi-group $\S$ of temporal
      supports with a unit element $*$ such that $s\circ*=*\circ s=s$
      for all $s\in\S$. A theory with no fundamental
\footnote{Of course, this does not rule out temporal characteristics
`emerging' in some coarse-grained sense. Indeed, any proper theory of
quantum gravity may well be of this type.}
temporal properties at all has $\S=\{*\}$.

\item There is a {\em null\/} history $0$ and a {\em
      unit\/} history $1$, both of whose support is $*$. These
      can be combined with any history filter $\a$ to give
      \beq
           \a\circ 1=1\circ\a=\a
      \eeq
      and
      \beq
           \a\circ0=0\circ\a=0.
      \eeq

\item The function $\s:\U\map\S$ that maps a history filter to
      its support is a homomorphism of the partial semi-groups. It may
      be profitable to factorise it as a tower of semi-groups as in
      \eq{tower}.
\ee

\subsection{Nuclear supports and nuclear history filters}
In standard quantum theory, any history filter $\a:=\h{\a}$ with
support $\{t_1,t_2,\ldots,t_n\}$ can be written as the composition
\beq
    \a=\a_{t_1}\circ\a_{t_2}\circ\ldots\circ\a_{t_n}\label{Decomp_a}
\eeq
in which the single-time proposition $\a_{t_i}$, $i=1,2,\ldots,n$  is
regarded as a history filter whose temporal support is the singleton set
$\{t_i\}$.

    In a general history theory it is important to know when a history
filter can be decomposed into the product of other filters in an
analogous way. The following general definitions are useful for this
purpose.
\be
\item A history filter $\a\in\U$ is {\em nuclear\/} if it cannot be
      written in the form $\a=\a_1\circ\a_2$ with both constituent
      history filters $\a_1,\a_2\in\U$ being different from the unit
      history.

\item A support $s\in\S$ is {\em nuclear\/} if it cannot be written in
      the form $s=s_1\circ s_2$ with both constituent supports
      $s_1,s_2\in\S$ being different from the unit support $*$.

\item A decomposition of $\a$ of the form
      $\a=\a_1\circ\a_2\circ\ldots\a_N$ is {\em irreducible\/} if the
      constituent history filters $\a_i$, $i=1\ldots N$, are all nuclear.
\ee

    Note that, in the decomposition \eq{Decomp_a} in standard quantum
theory, the constituents $\a_{t_i}$ are nuclear histories in the sense
above, and $\{t_i\}$ is a nuclear support. Thus, in a general history
theory, a nuclear support can be viewed as a (potentially
far-reaching)  analogue of a `point of time' in standard Hamiltonian
quantum theory; in particular, it admits no further temporal-type
subdivisions. Similarly, a nuclear history filter is a general
analogue of a single-time proposition.

\subsection{The partial ordering and `and' operations on $\U$}
If the set of all propositions about the histories of a system is
to take on a coherent algebraic form, it might be expected to be
equipped with a lattice structure that is analogous to the lattice
$\L$ of single-time propositions in standard quantum theory, and which
interfaces appropriately with the semi-group structure of the space of
supports. Let us analyse the separate operations in turn for the special
case of the history version of standard quantum logic.

\noindent
{\em The partial ordering on $\U$.\  }
\noindent
In standard quantum theory, a partial ordering can be defined on the
space of history filters by saying that a history filter
$\a\in\U(\L):={\cal F}(\T,\L)$ is {\em finer\/} than a history filter
$\b$ if, for all time points $t\in\T$, we have $\a_t<\b_t$ where $<$
denotes the usual partial-order operation on the lattice $\L$ of
fixed-time propositions. We expect such fine- (and coarse-) graining
relations to be universal, and therefore postulate that
in any general history theory:
\be
    \item the set $\U$ of history filters is a partially ordered set;
    \item the null and unit filters satisfy $0<\a<1$ for all $\a\in\U$;
    \item if $\a\in\U$ can be decomposed as $\a=\a_1\circ\a_2$ then
          $\a<\a_1$ and $\a<\a_2$.
\ee
The third requirement is natural if we bear in mind that the
heuristic meaning of $\a<\b$ is that the proposition that `universe
$\a$ is realised' contains more information than the same statement
made about $\b$; in particular, the statement that `$\a_1$ {\em and
then} $\a_2$ is realised' contains more information than either of the
two statements `$\a_1$ is realised', or `$\a_2$ is realised', alone.

\noindent {\em The `and' operation on $\U$.\  }
\noindent
In standard quantum theory, a natural `and' operation on a pair of
history filters $\a:=\h{\a}$ and $\b:=\h{\b}$ in $\U(\L)$ can be
defined by
\beq
    \a\wedge\b:=\h{(\a\wedge\b)}
\eeq
where $(\a\wedge\b)_{t_i}:=\a_{t_i}\wedge\b_{t_i}$ is the `and'
operation on the lattice $\L$. In the case of a Hilbert space
representation of $\L$ as a lattice of projection operators $P(\H)$,
$\op\a_{t_i}\wedge\op\b_{t_i}$ is defined to be the operator that
projects onto the intersection of the ranges of $\op\a_{t_i}$ and
$\op\b_{t_i}$. Heuristically, this says that the system gets through
history filter $\a$ and history filter $\b$ if, and only if, at each
time point it passes the single-time projector in both
filters. For this reason we set $\a\wedge\b=0$ if there is some time point
$t_i$ such that $\a_{t_i}\wedge\b_{t_i}=0$.

    This definition can be extended readily to history filters with
different temporal supports by defining $\a\wedge\b\in{\cal F}(\T,\L)$
as
\beq
    (\a\wedge\b)_t:=\a_t\wedge\b_t {\rm \ for\ all\ } t\in\T\simeq\R
                                                \label{Def:a_wedge_b}
\eeq
for all $\a,\b\in \U(\L):={\cal F}(\T,\L)$. Of course, for most
$t$-values, a history filter is the unit proposition and, for any
single-time proposition $P$, $1\wedge P=P\wedge1=P$.

    It is clear from this definition that if $\b$ follows $\a$ then
$\a\circ\b=\a\wedge\b$, and this relation is one we expect to
hold in the general case. Thus the following axiom can be added
for a general history theory:
\be
    \item The set $\U$ of history filters is equipped with a binary,
associative operation called `and' (denoted $\wedge$) such that,
heuristically, to say that $\a\wedge\b$ is realised means that both
$\a$ and $\b$ are realised. This operation is commutative,\ie
$\a\wedge\b=\b\wedge\a$ for all $\a,\b\in\U$.

    \item This `and'-operation is compatible with the partial-ordering
$<$ so that $\U$ is a {\em meet semi-lattice\/}; \ie $\a\wedge\b$ is a
greatest lower bound for $\a$ and $\b$.

    \item If $\b$ follows $\a$ then $\a\circ\b=\a\wedge\b$.
\ee

\subsection{The space of history propositions}
The next stage is to give a meaning to the negation $\neg\a$ or
disjunction $\a\vee\b$ of history filters. These operations should
possess various familiar properties. For example,  the negation
operation should satisfy
\beq
        \neg(\neg\a)=\a
\eeq
and
\beq
        \neg1=0,{\rm\ and\ } \neg0=1
\eeq
where $0$ and $1$ are the null and unit history filters.

    Let us start with standard Hamiltonian quantum theory. Note first
that the definition \eq{Def:a_wedge_b} is actually the `and' part of
the weak direct product $\prod_{t\in\T}\L_t$ of copies of the lattice
$\L$ indexed by the elements of $\T$. This suggests that we might look
at the `not' and `or' operations in this product lattice, which
are defined analogously to \eq{Def:a_wedge_b} as
\beq
    (\neg\a)_t:=\neg(\a_t)
\eeq
and
\beq
    (\a\vee\b)_t:=\a_t\vee\b_t
\eeq
where $\a_t\vee\b_t$ is the usual `or' operation in the lattice $\L$.
In an operator realisation this is the projection onto the closure of
the linear span of the ranges of the projection operators $\op\a_t$
and $\op\b_t$; the negation operation is $\neg\op\a_t=\op1-\op\a_t$.

    These operations may seem natural but they are not the correct
ones for our purposes. For example, if $(a,b)$ and $(c,d)$ are two
history filters with the same support $\{t_1,t_2\}$, the definition
above sets $(a,b)\vee(c,d)$ equal to the history filter $(a\vee
c,b\vee d)$, which (na\"{\i}vely) is true if $a$ or $c$ is true at time
$t_1$, and then $b$ or $d$ is true at time $t_2$. In particular, this
would be true if $a$ is true at time $t_1$ and then $d$ is true at
time $t_2$, But this corresponds to the history filter $(a,d)$,
whereas what we want to say is that the statement that
`$(a,b)\vee(c,d)$ is realised' means that either the filter $(a,b)$ is
realised, or the filter $(c,d)$ is realised.

    Similarly, according to the above, $\neg(a,b)=(\neg a,\neg b)$
which is true if and only if both $a$ and $b$ are false, whereas what
we are looking for should be true if either $a$ is false or $b$ is
false.

    The implication is that the sought-for `or' and
`not' operations do not map history filters into history filters but
rather take their values in a larger space of all {\em history
propositions\/}, denoted $\UP(\L)$. Of course, this conclusion is not new:
for example, it is reflected in the fact that, in the standard
representation of $\a$ by $\op{C}_\a$, the history `$\a$ {\em or\/}
$\b$' for disjoint homogeneous histories $\a$ and $\b$ is represented
by $\op{C}_\a+\op{C}_\b$, which is generally {\em not\/} a product of
projection operators and hence is not a representation of a
homogeneous history; \ie there is no homogeneous history $\g$ such
that $\op{C}_\a+\op{C}_\b=\op{C}_\g$. Similarly, $\neg\a$ is
represented by $\op1-\op{C}_\a$, which is usually not equal to $\op{C}_\b$
for any homogeneous history $\b$.

    In summary, the axioms presented so far for the space $\U$ and its
support space $\S$ suggest that, in the general case, $\U$ should be
regarded as a subset of a larger space $\UP$ of all history
propositions which has the structure of an orthocomplemented lattice.
Notions such as disjointness can then be defined in the usual way. For
example, $\a$ is disjoint from $\b$ is defined to mean $\a<\neg\b$. We
would like to know what other structure $\UP$ should have, in
particular what type of orthocomplemented lattice it is (for example:
orthomodular, atomic, with the covering property, \etc). However, it
would be much easier to speculate on such matters if we had a
mathematical model for the $\UP$-space of standard quantum theory with
the property that all the propositions are represented by projection
operators. The next step therefore is to find such a model.

\section{An `HPO' Version of Standard Quantum Theory}
\label{Sec:HPO}
\subsection{A restatement of the main problem}
In standard quantum logic, a history filter $\a$ is a set $\h{\a}$ of
single-time propositions. In a Hilbert space realisation, it is a
collection  $\oph{\a}$ of projection operators that can be regarded as
an element of the direct sum $\oplus_{t\in\{t_1,\ldots,t_n\}}B(\H)_t$
of $n$ copies of the algebra $B(\H)$ of bounded operators on the
Hilbert space $\H$. As such, it is itself a projection operator, as is
to be desired for a representative of a proposition (the proposition
is that the history $\a$ of the universe is `realised'). Furthermore,
the `and' and partial-ordering operations defined in this product
lattice have the desired physical interpretation.

    However, as emphasised above, this is not the case for the `or'
and `not' operations, and therefore the theory must be extended from
$\U$ to a bigger space $\UP$ of history propositions. In the standard
approach to history quantum theory, this problem is solved by noting
that, in the computation of the decoherence functional, only the
product operator  $\op{C}_\a:=
\op\a_{t_n}(t_n)\op\a_{t_{n-1}}(t_{n-1})\ldots\op\a_{t_1}(t_1)$ is
involved and, furthermore, the putative join $\a\vee\b$ of two
disjoint homogeneous histories can be represented in this sense as the
sum $\op{C}_\a+\op{C}_\b$. From this perspective, it might seem
tempting to work always with the $\op{C}_\a$ operator representative
\footnote{The operator $\op{C}_\a$ depends explicitly on the dynamics of
the system because it uses the {\em Heisenberg\/}-picture
operators $\op\a_{t_i}(t_i)$. A dynamics-independent version
is the simple product
$\op\a_{t_n}\op\a_{t_{n-1}}\ldots\op\a_{t_1}$ of the original
Schr\"odinger picture operators $\op\a_{t_i}$.}
of any homogeneous history $\a$ and to perform manipulations on these
objects alone. For example,  if $\b$ follows $\a$ the representative
$C$-operator for $\a\circ\b$ is
$\op{C}_{\a\circ\b}=\op{C}_\b\op{C}_\a$, and if $\a$ and $\b$ are
disjoint the representative operator is
$\op{C}_{\a\vee\b}:=\op{C}_\a+\op{C}_\b$. However, for our purposes,
such a procedure has several defects. In particular:
\bi
\item the map $\a:=\h{\a}\mapsto \op{C}_\a$ is many-to-one and
      therefore much information about $\a$ is lost in the process;

\item $\op{C}_\a$ is not a projection operator, and therefore it is not
      part of the propositional lattice associated with the Hilbert space
      $\H$ on which it is defined. This makes it difficult to know
      what the $C$-representative is of, for example, $\a\vee\b$ when
      $\a$ and $\b$ are {\em not\/} disjoint.
\ei

    The intention now is to construct a different operator
representation of standard quantum theory in which  every history
proposition {\em is\/} represented by a genuine projection operator,
thus enabling the whole of $\UP$ to be identified with the projection
lattice of some new Hilbert space. For reasons that will become clear
later, we are obliged at this point to restrict our attention to the
special case where $\L=P(\H)$, \ie we must work with a normal
Hilbert-space based quantum system rather than a general lattice $\L$.
However, the result still strongly reinforces the supposition that,
for a {\em general\/} history theory, the space $\UP$ is an
ortho-complemented lattice that might be representable by projection
operators. I shall call any such mathematical structure an {\em HPO
theory\/}, where HPO stands for `history projection operator'.

\subsection{The HPO-theory for standard quantum theory with a fixed
temporal support}
\label{SSec:HPO_fixedTS}
If $\P$ and $\Q$ are projection operators, the
product $\P\Q$ generally fails to be so because, unless
$[\P,\Q]=0$, it is neither hermitian nor idempotent. This is
why $\op{C}_\a$ is generally not a projection operator.

    However, consider instead the {\em tensor\/} product
$\P\otimes\Q$. The product of operators $\op{A}\otimes\op{B}$ and
$\op{C}\otimes\op{D}$ is defined as
$(\op{A}\otimes\op{B})(\op{C}\otimes
\op{D}):=\op{A}\op{C}\otimes\op{B}\op{D}$, while the adjoint operation
is $(\op{A}\otimes\op{B})^{\dag}:=\op{A}^{\dag}\otimes\op{B}^{\dag}$.
It follows at once that $\P\otimes\Q$ {\em is\/} a projection
operator, and is hence a candidate to represent the two-time
homogeneous history $(\P,\Q)$. More generally, if we consider the set
$\U_{\{t_1,\ldots,t_n\}}$ of all homogeneous histories with (for the moment)
a fixed support $\{t_1,t_2,\ldots,t_n\}$, let us represent any such
$\a=\oph{\a}$ with the tensor product
\beq
    \th\oph{\a}:=\tp{\a}                          \label{Def:th}
\eeq
which acts on the tensor-product space $\otimes_{t\in\{t_1,\ldots,
t_n\}}\H_t$ of $n$ copies of $\H$.

    That the tensor product appears in a natural way can be seen from
the following observation. The homogeneous history $\a:=\oph{\a}$ belongs
to the direct sum $\oplus_{t\in\{t_1,\ldots t_n\}}B(\H)_t$ of
$n$-copies of the operator algebra $B(\H)$. In constructing the
decoherence functional, the map
\beq
    \oph{\a}\mapsto \tr\big(\op\a_{t_n}(t_n)\op\a_{t_{n-1}}(t_{n-1})
        \ldots\op\a_{t_1}(t_1)\op{B}\big)
\eeq
is {\em multilinear\/} with respect to the vector space structure of
$\oplus_{t\in\{t_1,\ldots t_n\}}B(\H)_t$ for any $\op{B}\in B(\H)$.
However, the fundamental property of the tensor product of a finite
collection of vector spaces $V_1,V_2,\ldots, V_n$ is that any
multilinear map $\mu:V_1\times V_2\times\ldots\times V_n\map W$ to a
vector space $W$ factorises uniquely through the tensor product to give
the chain of maps
\beq
    V_1\times V_2\times\ldots\times V_n\buildrel\th\over\rightarrow
        V_1\otimes V_2\otimes\ldots\otimes V_n
            \buildrel\mu'\over\rightarrow W.
\eeq
Hence the map from $\a=\oph{\a}$ to $\tp{\a}$ arises naturally in
the histories approach to standard quantum theory.

    The map $\th$ from $\oplus_{t\in\{t_1,\ldots,t_n\}}B(\H)_t \map
\otimes_{t\in\{t_1,\ldots,t_n\}}B(\H)_t$ is many-to-one since, for
example, $(\l \op{A})\otimes(\l^{-1}\op{B})=\op{A}\otimes\op{B}$ for
all $\l\in\C$, $\l\ne 0$. However, when restricted to the subspace
$\prod_{t\in\{t_1,\ldots,t_n\}}P(\H)_t\subset
\oplus_{t\in\{t_1,\ldots,t_n\}}B(\H)_t$ of {\em projection\/}
operators, the map is {\em one\/}-to-one, essentially because if $\P$
is a projection operator, $\l\P$ ($\l\ne0$) is such only for $\l=1$.
Thus the first observation is that:
\bi
\item unlike the standard representation with $\op{C}_\a$, no
      information about the homogeneous history $\oph{\a}$ is lost by
      representing it with the tensor product $\tp{\a}$;

\item unlike $\op{C}_\a$, the operator $\tp{\a}$ is hermitian and
      idempotent, \ie it {\em is\/} a projection operator.
\ei

    The key observation is now the following. If $\P$ and $\Q$ are
projection operators on $\H$ then $\P\otimes\Q$ is a projection
operator on $\H\otimes\H$. Furthermore, if $\P$ and $\Q$ are disjoint
(\ie if $\P\Q=\Q\P=0$) then $\P+\Q$ is a projection operator on $\H$.
Now suppose that $\P$ and $\Q$ are disjoint projection operators on
$\H$, as are the pair $\op{R}$ and $\op{S}$; hence
$\P\Q=\Q\P=0=\op{R}\op{S}=\op{S}\op{R}$. Then
$(\P\otimes\op{R})(\Q\otimes
\op{S})=\P\Q\otimes\op{R}\op{S}=0=(\Q\otimes\op{S})(\P\otimes\op{R})$
and hence $\P\otimes\op{R}$ and $\Q\otimes\op{S}$ are disjoint
projection operators on $\H\otimes\H$ (actually, it is sufficient that
just {\em one\/} of the pairs $(\P,\Q)$ and $(\op{R},\op{S})$ is
disjoint). It follows that their {\em sum\/}
$\P\otimes\op{R}+\Q\otimes\op{S}$ is also a projection operator on
$\H\otimes\H$. In fact, using the lattice structure on
$P(\H\otimes\H)$ we get
\beq
        \P\otimes\op{R}\vee\Q\otimes\op{S}=\P\otimes\op{R}+\Q\otimes\op{S}
\eeq
which looks remarkably like the $\op{C}_\a+\op{C}_\b$ construction we are
trying to emulate.

    What this example illustrates is that not all projection operators
on $\H\otimes\H$ are simply tensor products of projection operators on
$\H$. Those that are, are normally called {\em homogeneous\/}; the
remainder are {\em inhomogeneous\/} and include certain sums of
homogeneous operators.

    The general strategy is now clear. With the aid of the map $\th$,
the operator representation $\prod_{t\in\{t_1,\ldots t_n\}}P(\H)_t$ of
the space of homogeneous histories $\U_{\{t_1,\ldots,t_n\}}$ with
temporal support $\{t_1,t_2,\ldots,t_n\}$ is embedded in the space
$P(\otimes_{t\in\{t_1,\ldots,t_n\}}\H_t)$ of projection operators on
the Hilbert space $\otimes_{t\in\{t_1,\ldots,t_n\}}\H_t$. The space
$P(\otimes_{t\in\{t_1,\ldots,t_n\}}\H_t)$ carries the usual lattice
structure of projection operators and is therefore a natural model for
the space $\UP_{\{t_1,\ldots,t_n\}}$ of history propositions based on
homogeneous histories with support $\{t_1,t_2,\ldots, t_n\}$. In this
model, history filters/homogeneous histories are represented by
homogeneous projectors, and a general history proposition is
represented by an inhomogeneous projector. This  explains why the
collection $\oph{\a}$ was refered to earlier as a `homogeneous'
history.

    In this approach, decoherence functionals will be computed
with the aid of the map $D:\otimes_{t\in\{t_1,\ldots,t_n\}}B(\H)\map
B(\H)$ defined by
\beq
    D(\op{A}_1\otimes\op{A}_2\otimes\ldots\op{A}_n):=
    \op{A}_n(t_n)\op{A}_{n-1}(t_{n-1})\ldots\op{A}_1(t_1) \label{Def:D}
\eeq
on homogeneous operators and then extended by linearity. Thus, on a
homogeneous history $\a\in\U_{\{t_1,\ldots,t_n\}}$, the $C$-map
is defined by
\beq
    \op{C}_\a:=D(\th(\a))                            \label{Def:C_a=D}
\eeq
and then extended by linearity to the set of inhomogeneous histories
with support $\{t_1,t_2,\ldots,t_n\}$.

\subsection{The missing logical connectives}
The natural lattice structure on $P(\otimes_{t\in\{t_1,\ldots,
t_n\}}\H_t)$ can be used to define the missing negation and `or'
operations on $\UP$. For example, in a general lattice of projection
operators, the projection operator corresponding to the proposition
$\neg\P$ is $\op{1}-\P$. Hence, if $\op{R}$ and $\op{S}$ are projection
operators on the Hilbert space $\H$, the negation of the proposition
represented by the projection operator $\op{R}\otimes\op{S}$ on
$\H\otimes H$ is just
\beq
        \neg(\op{R}\otimes\op{S})=\op{1}-\op{R}\otimes\op{S}
\eeq
and so, using the $D$-map defined in \eq{Def:D}, we get
$D(\neg(\op{R}\otimes\op{S}))=\op{1}-\op{S}\op{R}$. Thus, for the
homogeneous history $\a\in\U_{\{t_1,\ldots,t_n\}}$, \eqs{Def:D}{Def:C_a=D}
give
\beq
    \op{C}_{\neg\a}=\op{1}-\op\a_{t_n}(t_n)\op\a_{t_{n-1}}(t_{n-1})
                    \ldots\op\a_{t_1}(t_1)
\eeq
in accord with the usual identification made in work on decohering
histories.

    If $\a$ and $\b$ are disjoint homogeneous histories with the same
support $\{t_1,t_2,\ldots,t_n\}$ then, according to the philosophy
above, the history proposition $\a\vee\b$ is represented by the sum
$\th(\a)+\th(\b)=\tp{\a}+\tp{\b}$, and so
\beq
    \op{C}_{\a\vee\b}:=D(\th(\a)\vee\th(\b))=D(\tp{\a}+\tp{\b})=
                \op{C}_\a+\op{C}_\b
\eeq
as required. However, we can now describe the $C$-representation
of the proposition $\a\vee\b$ when $\a$ and $\b$ are {\em not\/}
disjoint. It is simply the image under the map
$D:\otimes_{t\in\{t_1,\ldots t_n\}}B(\H)\map B(\H)$ of the projection
operator $\th(a)\vee\th(b)$, \ie of the operator on
$\otimes_{t\in\{t_1,\ldots t_n\}}\H$ that projects onto
the smallest closed subspace containing the images of the projection
operators $\th(\a)=\tp{\a}$ and $\th(\b)=\tp{\b}$.

    There are situations other than disjointness in which this
operator can be written in a simple form. For example, if $\P$ and
$\Q$ are projection operators that commute, but are not disjoint, then
it is a standard result that $\P\vee\Q=\P+\Q-\P\Q$. Now suppose that
$(\op\a_{t_1},\op\a_{t2})$ and $(\op\b_{t_1},\op\b_{t_2})$ are a pair
of homogeneous histories with support $\{t_1,t_2\}$ (two time points
for simplicity only) and which are such that
$[\op\a_{t_1},\op\b_{t_1}]=0=[\op\a_{t_2},\op\b_{t_2}]$. Then the
tensor-product projection operators $\op\a_{t_1}\otimes\op\a_{t_2}$
and $\op\b_{t_1}\otimes\op\b_{t_2}$ also commute, and hence
\beqa
    \op\a_{t_1}\otimes\op\a_{t_2}\vee\op\b_{t_1}\otimes\op\b_{t_2}
    &=\op\a_{t_1}\otimes\op\a_{t_2}+\op\b_{t_1}\otimes\op\b_{t_2}-
        (\op\a_{t_1}\otimes\op\a_{t_2})(\op\b_{t_1}\otimes\op\b_{t_2})
                                                \nonumber\\
    &=\op\a_{t_1}\otimes\op\a_{t_2}+\op\b_{t_1}\otimes\op\b_{t_2}-
        \op\a_{t_1}\op\b_{t_1}\otimes\op\a_{t_2}\op\b_{t_2}
\eeqa
so that
\beq
    \op{C}_{\a\vee\b}=\op\a_{t_2}(t_2)\op\a_{t_1}(t_1)+
                    \op\b_{t_2}(t_2)\op\b_{t_1}(t_1)
        -\op\a_{t_2}(t_2)\op\b_{t_2}(t_2)\op\a_{t_1}(t_1)\op\b_{t_1}(t_1)
\eeq
which, perhaps, is not such an obvious result.

    Note that the tensor product structure is well-adapted
to discuss the sequential-logic operations introduced by MittelStaedt
\cite{Mit77} and Stachow
\cite{Sta80}. For example, a simple, time-labelled, sequential
disjunction is of the form `proposition $P$ at time $t_1$ {\em or
then\/} proposition $Q$ at time $t_2$', denoted $P\sqcup Q$. In terms
of history filters, the proposition `$P$ at time $t_1$'
says nothing about time $t_2$ and is therefore represented by
the tensor product operator $\P\otimes\op1$. Similarly, the
proposition `$Q$ at time $t_2$', says nothing about $t_1$, and is
therefore represented by the operator $\op1\otimes\Q$. Thus, the
sequential disjunction $P\sqcup Q$ can be identified with the operator
$\P\otimes\op1\vee\op1\otimes\Q$.

    Now, even if $\P$ and $\Q$ do not commute, we still have $[\P\otimes
\op1,\op1\otimes\Q]=0$ and hence
\beq
    \P\otimes\op1\vee\op1\otimes\Q =\P\otimes\op1+\op1\otimes\Q -
         (\P\otimes\op1)(\op1\otimes\Q)=
            \P\otimes\op1+\op1\otimes\Q-\P\otimes\Q
\eeq
so that, under the $C$ map, we get
\beq
    \op{C}_{P\sqcup Q}=\P+\Q-\Q\P
\eeq
in agreement with Stachow's result.

    Similarly, the sequential conjunction `$P$ at time $t_1$ {\em and
then\/} $Q$ at time $t_2$' (denoted $P\sqcap Q$) is represented by
$\P\otimes\op1\wedge\op1\otimes\Q$ which is equal to $\P\otimes\Q$,\ie
the image under $\th$ of the history filter $(P,Q)$. Thus, in
the case of standard quantum theory, a sequential conjunction (with
temporal labels) is indeed the same thing as a history
filter/homogeneous history.

    Finally, it is now clear why we had to restrict our attention at
the beginning of this section to the special case $\L=P(\H)$. This is
because it is not possible to give a sensible definition of the tensor
product of a general lattice $\L$ with itself; a well-known problem in
standard quantum logic.

\subsection{The HPO-theory for general supports}
The theory developed above works well but, as it stands, it applies
only to homogeneous histories with a fixed temporal support
$\{t_1,t_2,\ldots,t_n\}$ and to those general history propositions
that can be derived from them. To incorporate arbitrary supports we
need to collect together the operator algebras $\otimes_{t\in
s}B(\H)_t$ for all supports $s\in\S$. The natural way of doing this is
to use an {\em infinite\/} tensor product of copies of $B(\H)$.

    Let $\Omega$ denote a family of unit vectors in the Cartesian
product $\prod_{t\in\T}\H_t$ of copies of $\H$ labelled by the time
values $t\in\T$; \ie $\Omega$ is a map from $\T$ to the unit sphere in
$\H$. Recall that the infinite tensor product
$\otimes^\Omega_{t\in\T}\H_t$ based on $\Omega$ is defined to be the set
of functions $v:\T\map\H$ with the property that $v(t)=\Omega(t)$ for all
but a finite number of $t$-values. The set of all such functions
is given the usual point-wise vector space structure by defining
$(av+bw)(t):=av(t)+bw(t)$, $a,b\in\C$, $v,w\in
\otimes^\Omega_{t\in\T}\H_t$, and the scalar product is
\beq
        <v,w>:=\prod_{t\in\T}<v(t),w(t)>_\H \label{Def:<>InfTP}
\eeq
where $<,>_\H$ is the inner product on the Hilbert space $\H$. The
right hand side of \eq{Def:<>InfTP} is well defined because only a
finite number of terms contribute to the product. It is a standard
result that the resulting space is a Hilbert space \cite{Gui72}.

    An infinite tensor product $\otimes^\Omega_{t\in\T}B(\H)_t$ of
operator algebras $B(\H)$ is naturally associated with this new vector
space. It is defined to be the weak closure (\ie the closure in the
weak operator topology) of the set of all functions from $\T$ to
$B(\H)$ that are equal to the unit operator for all but a finite set
of $t$-values. This definition is geared perfectly to accomodate
arbitrary temporal supports, and therefore the set of all projection
operators in $\otimes^\Omega_{t\in\T}B(\H)_t$ can be taken as a model
for the complete space $\UP$ of history propositions in a standard
Hilbert-space based, quantum theory. Elements of the subspace $\U$ of
history filters are identified with the homogeneous projection
operators in $\otimes^\Omega_{t\in\T}B(\H)_t$, \ie functions from $\T$
to $P(\H)$ that are equal to the unit operator for all but a finite
set of $t$-values.

    Note that use of the weak closure in forming the complete space
$\otimes^\Omega_{t\in\T}B(\H)_t$ provides a precise way of giving
meaning to a history with an infinite set of projection operators:
it is the weak limit of a convergent sequence of homogeneous
histories with finite support.

\subsection{Group actions on the HPO space}
\label{SSec:GrpHPO}
An interesting insight into the possible role of group theory in the
general history formalism can be obtained by looking at what happens
in the HPO form of standard quantum theory. Thus let
$R:\T\map B(\H)$ be any map whose value in $B(\H)$ is the
unit operator on $\H$ for all but the finite set of time values in
some support $s=\{t_1,t_2,\ldots,t_n\}$. Then
\beq
    R(s):=R_{t_1}\otimes R_{t_2}\otimes\ldots\otimes R_{t_n}
                                            \label{Def:R(s)}
\eeq
is a well-defined operator on the tensor product space
$\otimes^\Omega_{t\in\T}\H_t$. Now suppose we are given such a map
$s\mapsto R(s)$ for each $s\in\S$. Then it is a trivial consequence of
definition \eq{Def:R(s)} that if $s':=\{t'_1,t'_2,\ldots,t'_m\}$ is
another support that follows $s$ (\ie $t_n<t'_1$) then
\beq
    R(s)R(s')=R(s\circ s').
\eeq
Thus the map $s\mapsto R(s)$ is a representation of the semi-group
$\S$ on the HPO Hilbert space.

    Now suppose that each $R_{t_i}$ is a {\em unitary operator\/} on
$\H$. Then the associated operator $R(s)$ is a unitary operator on
$\otimes^\Omega_{t\in\T}\H_t$, and hence, for any history filter $\a$
(with arbitrary temporal support), the map
\beq
    \op\a_{t'_1}\otimes\op\a_{t'_2}\otimes\ldots\otimes\op\a_{t'_m} \mapsto
R(s)\op\a_{t'_1}\otimes\op\a_{t'_2}\otimes\ldots\otimes\op\a_{t'_m}
                                        R(s)^{-1}
\eeq
maps the homogeneous projector
$\op\a_{t'_1}\otimes\op\a_{t'_2}\otimes\ldots\otimes\op\a_{t'_m}$ into
another such. This extends to a map of inhomogeneous projectors,
and in such a way that the lattice structure of the history
propositions is preserved. Thus, in the unitary case, the map
$s\mapsto R(s)$ provides a representation of the semi-group $\S$
as automorphisms of the lattice of history propositions. This
suggests very strongly that, in the general case, an important role
will be played by representations of the semi-group of supports $\S$
in the group of automorphisms of the lattice $\UP$.

    Note that, in standard quantum theory, a very special example is
the choice $R_t:=e^{-itH/\hbar}$. In this case, the tensor product
$\op\a_{t_1}(t_1)\otimes\op\a_{t_2}(t_2)\otimes\ldots\otimes
\op\a_{t_n}(t_n)$ of Heisenberg-picture projectors can be obtained
from the product $\tp{\a}$ of Schr\"odinger operators by the transform
\beq
\op\a_{t_1}(t_1)\otimes\op\a_{t_2}(t_2)\otimes\ldots\otimes
    \op\a_{t_n}(t_n)=U(s)\tp{\a}U(s)^{-1}
\eeq
where $s:=\{t_1,t_2,\ldots,t_n\}$ and $U_t:=e^{-itH/\hbar}$. This
suggests that, in the general case, if an HPO representation can be
found situations may arise in which a quasi-temporal type of
dynamical evolution, together with an associated Heisenberg picture,
can be achieved with the aid of a special operator representation of
$\S$.

\subsection{The general axioms for history propositions}
Motivated by the above we can now give a provisional form of the
expanded set of history axioms towards which we have been working.
These should be viewed as a more detailed version of the original
Gell-Mann and Hartle axioms and, hopefully, could be used as a basis
for a general analysis of quantum gravity theories in which the
standard notions of continuous time and space need not be fundamental
ingredients. The general axioms and definitions are as follows.

\noindent
H1. {\em The space of history filters.}\ \
\noindent
The fundamental ingredient in a theory of histories is a space $\U$ of {\em
history filters\/}, or {\em possible universes\/}. This space
has the following structure.
\be
\item $\U$ is a partially-ordered set with a {\em unit\/}
      history filter $1$ and a {\em null\/} history filter $0$
      such that $0<\a<1$ for all $\a\in\U$;

\item $\U$ has a meet operation $\wedge$ which combines with the
      partial order $<$ to form a meet semi-lattice with unit $1$ so
      that $1\wedge\a=\a$ for all $\a\in\U$. The null history is
      absorptive in the sense that $0\wedge\a=0$ for all $\a\in\U$.

\item $\U$ is a partial semi-group with composition law
      denoted $\circ$. If $\a,\b\in\U$ can be combined to give
      $\a\circ\b\in\U$ we say that $\b$ {\em follows\/} $\a$, or
      $\a$ {\em preceeds\/} $\b$, and write $\a\lhd\b$. The $\circ$ and
      $\wedge$ laws are compatible in the sense that if $\a\circ\b$ is
      defined then it is equal to $\a\wedge\b$.

\item The null and unit histories can always be combined with any
      history filter $\a$ to give
      \beqa
        \a\circ 1&=&1\circ\a=\a         \\
        \a\circ 0&=&0\circ\a=0.
      \eeqa
\ee

\noindent
H2. {\em The space of temporal supports.}\ \
\noindent
Any quasi-temporal properties of the system are encoded in a partial
semi-group $\S$ of supports with unit $*$. The support space has the
following properties.
\be
\item There is a semi-group homomorphism $\s:\U\map\S$ that assigns a
      support to each history filter. The support of $0$ and $1$ is
      defined to be $*\in\S$.

\item {A history filter $\a$ is {\em nuclear\/} if it has no non-trivial
      decomposition of the form $\a=\b\circ\g$, $\b,\g\in\U$; a temporal
      support $s$ is {\em nuclear\/} if it has no non-trivial decomposition
      of the form $s=s_1\circ s_2$, $s_1,s_2\in\S$. Nuclear supports are the
      analogue of points of time; nuclear history filters are the
      analogue of single-time propositions.

      A decomposition of $\a\in\U$ as
      $\a=\a^1\circ\a^2\circ\ldots\circ\a^N$ is {\em irreducible\/} if
      the constituent history filters $\a^i\in\U$, $i=1,2,\ldots,N$ are
      all nuclear.
      }
\item A {\em resolution\/} of the semi-group homomorphism $\s:\U\map\S$
      is a chain of semigroups $\U_{i}$ and semi-group homomorphisms
      $\s_i$ so that $\s$ factors as the composition
      \beq
            \U{\buildrel\s_0\over\rightarrow}\U_1
              {\buildrel\s_1\over\rightarrow}\U_2
              {\buildrel\s_2\over\rightarrow}\ldots
              {\buildrel\s_{k-1}\over\rightarrow}\U_k
              {\buildrel\s_k\over\rightarrow}\S.
      \eeq
\ee

\noindent
H3. {\em The space of history propositions.}\ \
\noindent
The space $\U$ of history filters is embedded in a larger space $\UP$
of {\em history propositions\/}. This space is a
ortho-complemented lattice with a structure that is consistent with
the semi-lattice structure on the subspace $\U$. One may also require
the lattice to be countably complete, or even complete, depending on
its cardinality. In addition:
\be
\item The space $\UP$ can be generated from $\U$ by the
      application of a finite (or, perhaps, countably infinite) number
      of $\neg$, $\vee$ and $\wedge$ lattice operations. This captures
      the idea that elements of $\UP$ represent propositions `about'
      history filters (\ie about possible universes).

\item An important role may be played by representations of the
      partial semi-group $\S$ in the automorphism group of the lattice
      $\UP$.

\item Any representation of the ortho-complemented lattice $\UP$ by
      projection operators on a Hilbert space is called an {\em
      HPO quantisation\/} of the system. In such a quantisation,
      some of the automorphisms mentioned above might be implemented
      with unitary transformations.

\item Two history propositions $\a$ and $\b$ are said to be {\em
      disjoint\/}, denoted $\a\perp\b$, if $\a<\neg\b$. A set of
      history propositions $\{\a^1,\a^2,\ldots,\a^N\}$ is {\em exclusive\/}
      if its elements are pairwise disjoint. It is {\em exhaustive\/} (or
      {\em complete\/}) if $\a^1\vee\a^2\vee\ldots\vee\a^N=1$.
      Countable sets (\ie with $N=\infty$) are permitted where appropriate.
\ee

\noindent
H4. {\em The space of decoherence functionals.}\ \
\noindent
A {\em decoherence functional\/} is a complex-valued map of pairs of
history propositions; the set of all such maps is denoted $\cal D$.
Any decoherence functional has the following properties:
\be
\item The `inner-product' type conditions:
      \bi
      \item {\em Hermiticity\/}:\ \ $d(\a,\b)=d(\b,\a)^*$ for all
            $\a,\b\in\UP$.
      \item {\em Positivity\/}:\ \ $d(\a,\a)\ge0$ for all $\a\in\UP$.
      \item {\em Null triviality\/}:\ \ $d(0,\a)=0$ for all $\a\in\UP$.
      \ei

\item Conditions related to the potential probabilistic interpretation:
      \bi
      \item {\em Additivity\/}:\ \ if $\a\perp\b$ are general history
            propositions then, for all $\g\in\UP$,
            $d(\a\vee\b,\g)=d(\a,\g)+d(\b,\g)$.
      \item {\em Normalisation\/}
\footnote{This condition, plus additivity, implies that
if $\a^1,\a^2,\ldots,\a^N$ and $\b^1,\b^2,\ldots,\b^N$ are two
complete sets of history propositions then
$\sum_{i=1}^N\sum_{j=1}^Md(\a^i,\b^j)=1$. In the infinite case one may
wish to extend this to allow both $N$ and $M$ to be infinite.} : $d(1,1)=1$

      \ei

\item There may be a natural topology on $\UP$ such that each
decoherence functional $d\in{\cal D}$ is a {\em continuous\/}
\footnote{Such a condition holds in standard quantum theory because
of the result that $\op{A}\mapsto \tr(\op{A}\op{B})$ is a weakly
continuous function on bounded subsets of $B(\H)$ for each trace-class
operator $\op{B}$.}
function of its arguments.
\ee

\noindent
H5. The physical interpretation of these axioms is the same as that
of the Gell-Mann and Hartle axioms, \ie the diagonal element
$d(\a,\a)$ is interpreted as the probability of the history
proposition $\a$ being `true' when $\a$ is part of a consistent set.
If this is not the case, no direct physical meaning is ascribed to the
real number $d(\a,\a)$.

\smallskip
    The axioms above might be tightened or relaxed in a variety of
ways; certainly one would not wish to be too pedantic at this stage of
the development. For example, Dowker and Halliwell \cite{DH92} have
pointed out that the decoherence functional \eq{Def:d_rho} in standard
Hamiltonian quantum theory obeys the Schwarz-type inequality
\beq
    |d(\a,\b)|\le d(\a,\a)^{1\over2}d(\b,\b)^{1\over2}\label{SI_d}
\eeq
for all history filters $\a$ and $\b$. This might be a good candidate
for inclusion as an extra axiom in the case of a general history
theory. Note that it renders the `null triviality' axiom redundant.

    In normal quantum theory, the lattice $\L$ of propositions is
not only complete and ortho-complemented: it is also orthomodular,
atomic and with the covering property. Each of these properties has a
direct physical implication, and it is therefore a matter of some
interest to explore the extent to which they should also be imposed on
the lattice $\UP$ of history theory.

    In this context it is appropriate to recall the powerful
representation theorems that assert that any complete, orthomodular
atomic lattice with the covering property (and of length $\ge4$) can
be realised as the lattice of closed subspaces of a vector space $V$
equipped with a hermitian form \cite{Pir76}, \cite{BC81}. If the field
over which $V$ is defined is the real, complex or quaternionic numbers
then, under a minor continuity requirement on the involution operator
(complex conjugation in the complex case), $V$ is a Hilbert space.  It
follows that if the space $\UP$ of history propositions is a lattice
of this type, it can be realised as the lattice $P({\cal V})$ of
projection operators on some Hilbert space $\cal V$, \ie an
`HPO-theory' necessarily exists. Of course, the HPO-Hilbert space
$\cal V$ need not be anything like the tensor-product structure
defined above for standard quantum theory; clearly, this will depend
on the properties of the temporal supports. On the other hand, if
either the orthomodular or the atomicity condition is absent the
possibility arises of finding mathematical models that are not
connected with Hilbert spaces at all. This is an intriguing topic for
further research.

\subsection{The non-applicability of Gleason's theorem to the
HPO-theory}
We have seen that all propositions about histories in standard
Hamiltonian quantum theory can be represented by projection operators
on the Hilbert space $\cal V$ of the HPO-theory. It is instructive to see
how Gleason's theorem \cite{Gle57} looks in this context.

    Let $\mu$ be any state on the lattice $P(\H)$ of projection
operators on a Hilbert space $\H$, \ie $\mu$ is a real-valued function
on $P(\H)$ that satisfies the axioms \eqs{Def:state_1}{Def:state_3}.
Then Gleason showed that, if Dim$(\H)>3$, there exists a density
matrix $\r_\mu$ on $\H$ such that, for all $\a\in P(\H)$,
\beq
    \mu(\a)=\tr(\r_\mu\a)                   \label{Gleason}
\eeq
which, of course, is the familiar expression for the
probability in standard quantum theory.

    Gleason's theorem is important because it sharply limits the ways
in which the quantum-logical structure contained in the pair
$(\L,{\cal R})$ can be represented in concrete mathematical terms.
More precisely, if the lattice of propositions $\L$ is postulated to
be the projection lattice $P(\H)$ of some Hilbert space $\H$, then
Gleason's theorem shows that the states must necessarily be of the
standard form \eq{Gleason}. Thus, to construct a non-standard type of
quantum theory in which states are still additive on
all pairs of disjoint propositions it is necessary to start with a
model for $\L$ that is {\em not\/} the projection lattice of a Hilbert
space.

    Consider now the case where $\cal V$ is the Hilbert space of a
general HPO theory, \ie the set of history propositions
$\UP$ is represented by the lattice $P({\cal V})$ of projection
operators on $\cal V$. Each decoherence functional $d\in{\cal D}$
defines a real-valued map $\mu_d$ on $\UP$ by $\mu_d(\a):=d(\a,\a)$.
This certainly satisfies the first two requirements
\eqs{Def:state_1}{Def:state_2} of an abstract state and therefore, if
this map was additive on {\em all\/} complete sets of history
propositions, Gleason's theorem would imply that it is
necessarily  given by a density matrix acting on $\cal V$ via
\eq{Gleason}. Under these circumstances, it would be rather difficult
to tell the difference between a history theory and a normal
(single-time) theory.

    However, the central idea behind the notion of consistent histories is
precisely that the decoherence functional is {\em not\/} additive in
this sense, \ie even if the histories $\a$ and $\b$ are disjoint,
$\mu_d(\a\vee\b)=d(\a\vee\b,\a\vee\b)$ may not equal
$\mu_d(\a)+\mu_d(\b)=d(\a,\a)+d(\b,\b)$. In fact, this is generally
the case since additivity applies only to the special subsets of
`consistent' histories. Thus Gleason's theorem is not applicable, and
so it is feasible to seek decoherence functionals on $P({\cal V})$ that are
something other than density matrices on $\cal V$. Indeed, this feature is
exploited already in standard quantum theory since the decoherence
functional $d(\a,\b):=\tr(\op{C}_\a\op{\r}\op{C}_\b^{\dag})$ cannot be
written in terms of a density matrix on the HPO-Hilbert space
$\otimes_{t\in\T}^\Omega\H_t$.

    This suggests rather strongly that interesting general history
theories might be constructed by postulating the space of history
propositions $\UP$ to be the projection lattice $P({\cal V})$ of some
Hilbert space $\cal V$, and then to look for decoherence functionals
whose diagonal elements are not density matrices on $\cal V$. For this
reason it is of considerable importance to see what can be said in
general about the space of decoherence functionals associated with the
projection lattice of a Hilbert space. This is another topic for
future research.

\section{Quantum Field Theory in a Curved Space-time}
\label{Sec:QFT}
The application of consistent histories to quantum field theory was
discussed by Blencowe \cite{Ble91} in the case of a
scalar field $\f$ propagating on a four-dimensional manifold $\M$ with
a globally-hyperbolic metric $\g$.  Global hyperbolicity guarantees
the existence of a global time function that can be used to foliate
$\M$ into a one-parameter family, $t\mapsto\Sigma_t$, of spacelike
hypersurfaces $\Sigma_t$ of $\M$.  Blencowe defined a history to be a
collection of propositions about the values of a sequence $A_{t_i}$,
$i=1,2,\ldots,n$ of observables where $A_{t_i}$ is localised in a
bounded open subset $O_{t_i}$ of $\Sigma_{t_i}$; these regions can be
time-ordered by requiring $t_1<t_2<\ldots<t_n$. Thus the global
hyperbolicity of $\g$ is used to reduce the problem to something
resembling standard Hamiltonian quantum theory.

    A more space-time oriented approach was suggested by Hartle
\cite{Har91a} who emphasised the importance of observables regarded as
averages of field variables over space-time regions. The causal
structure involved is related to ideas of Sorkin \cite{Sor93a} about
state-vector reduction in relativistic quantum theory. Other
significant work in this area is the general study by Mittelstaedt,
and by Mittelstaedt and Stachow \cite{Mit83a,Mit83b,MS83,Mit85}, of
`relativistic quantum logic': a quantum-logical version of the ideas
of space-time localised observables that are much used in $C^*$-algebra
approaches to quantum field theory.

    My aim here is to show briefly how the ideas of Hartle and Sorkin
lead naturally to a nice example of a `temporal support' in a
quasi-temporal situation. A lot more could be said about the quantum
field theoretical aspects of the theory, but this is deferred to
another occasion

    We start with a pair $(\M,\g)$ where the Lorentzian metric
$\g$ is not required to be globally hyperbolic (this is one part of the
`quasi-temporal' aspect of the system). The basic propositions in the
theory will be of the form that the smeared field $\f(f)$ lies in some
subset $I$ of $\R$ where $f$ is smooth test function on $\M$ (this
proposition will be denoted $P(f,I)$). The `history' concept comes in
via the idea of looking at a collection of such propositions
in which the supports of the test-functions concerned are localised in
space-time regions that are causally related in some way.

    To facilitate this idea, let us say that a subset $A\subset\M$ {\em
preceeds\/} another subset $B\subset\M$, denoted $A\prec B$, if
\beq
        A\bigcap B=\emptyset,           \label{Def:AprecB_1}
\eeq
\beq
        J^+(A)\bigcap B\ne\emptyset     \label{Def:AprecB_2}
\eeq
and
\beq
        J^+(B)\bigcap A=\emptyset       \label{Def:AprecB_3}
\eeq
where the {\em chronological future\/} $J^+(A)$ of $A$ is defined as
usual to be the set of points in $\M$ that can be reached from $A$ by
future-directed, non space-like curves. The purpose of
\eq{Def:AprecB_3} is to avoid a situation like that shown in Figure 2 where
$A$ and $B$ can each causally affect the other.

    As emphasised by Sorkin \cite{Sor93a}, the relation $\prec$ is
{\em not\/} a partial ordering; in particular, $A\prec B$ and $B\prec
C$ does not necessarily imply that $A\prec C$. This is because the
points in $B$ that causally influence $C$ (\ie $\{b\in B\mid
J^+(b)\bigcap C\ne\emptyset\}$) may not lie in that part of $B$
($B\bigcap J^+(A)$) that intersects the causal future $J^+(A)$ of $A$.
This is illustrated in Figure 3.

    The basic idea of the history formalism for this theory is that an
`elementary' history filter is a set of propositions $P(f_i,I_i)$,
$i=1,2,\ldots,n$ in which the support of each $f_i$ is {\em compact\/}
(so that the proposition is localised in the space-time $\M$) and {\em
connected\/}. Propositions involving regions that are disconnected can
be built up from propositions localised in connected regions.
This motivates the following definitions.
\be
\item An open subset $A\subset\M$ is a {\em basic\/} region of $\M$
      if $A$ is connected and has a compact closure.

\item A {\em temporal support\/} is a collection $s=\{O_1,O_2,\ldots,O_n\}$
      of basic regions $O_i$ whose closures are non-intersecting,
      and which are such that, for each pair $O_i,O_j\in s$, either (i)
      $O_i\prec O_j$, or (ii) $O_j\prec O_i$, or (iii) $O_i$ and $O_j$
      are space-like separated.

\item {A temporal support $s':=\{O'_1,O'_2,\ldots,O'_m\}$ is said to
      {\em follow\/}
      another such $s:=\{O_1,O_2,\ldots,O_n\}$, denoted $s\lhd s'$, if
      (i) the closures of the basic regions in $s$ and $s'$ are pairwise
      disjoint, and (ii) $\bigcup_{i=1}^n O_i \prec\bigcup_{j=1}^m O'_j$.

      If $s\lhd s'$, a semi-group combination law is defined as
      \beq
        s\circ s':=\{O_1,O_2,\ldots,O_n,O'_1,O'_2,\ldots,O'_m\}
      \eeq
      }

\item {A {\em history filter\/} is a collection of propositions
      $P(f_1,I_1),P(f_2,I_2)\ldots,P(f_n,I_n)$ where the support of
      the test function $f_i$, $i=1,\ldots,n$ is the closure of the
      element $O_i$ of a temporal support $\{O_1,O_2,\ldots,O_n\}$.

      If $\a,\b$ are history filters with $\s(\a)\lhd\s(\b)$, then
      $\a\circ\b$ is defined as the union of the collections of
      propositions contained in $\a$ and in $\b$.
      }
\ee

    Note that if a temporal support $s$ is nuclear, (\ie if it cannot be
written in the form $s_1\circ s_2$) then no $O_i\in s$ can proceed or
follow any other $O_j\in s$. Thus each $O_i$ in $s$ must be {\em
space-like\/} separated from every other one. Thus the role of a
`time-point' is played by any finite collection of disjoint basic
regions that are space-like separated from each other. It is clear
that every temporal support $s$ can be written in the form $s=s_1\circ
s_2\circ\ldots\circ s_N$ where $s_i$, $i=1\ldots N$ are nuclear sets. This
means that the time space $\cal T$ (defined to be the collection of
all possible nuclear temporal supports) can serve the same role here
as did the real numbers in our earlier discussion of standard Hamiltonian
quantum theory. In particular, an HPO theory can be
constructed in which the most general history proposition is
represented as a projection operator on the Hilbert space
$\otimes^\Omega_{\tau\in{\cal T}}\H_\tau$.

    The computation of decoherence functionals follows the same line
as that discussed by Hartle \cite{Har91a} and by Sorkin \cite{Sor93a}.
Specifically, in constructing the $C$-representation of a history
filter, if test-functions $f_i$ and $f_j$ have supports associated
with basic regions $O_i$ and $O_j$ with $O_i\prec O_j$, then the
projection operators representing the basic propositions $P(f_i,I_i)$
and $P(f_j,I_j)$ must appear in the order
$\op{P}(f_j,I_j)\op{P}(f_i,I_i)$. If $O_i$ and $O_j$ are spacelike
separated then no order for the operators $\op{P}(f_i,I_i)$ and
$\op{P}(f_j,I_j)$ can be specified, which is consistent only if they
{\em commute\/}. So this is the point at which microcausality enters
the history version of quantum field theory in a curved space-time.

    Specific quantum field theoretic problems will not be addressed
here but it should be noted that the approach discussed above does not
help with the difficult problem of deciding how concrete theories
should actually be constructed. The basic problem is that, viewed
canonically, a quantum field theory admits many unitarily inequivalent
representations of the canonical commutation relations, and some way
is needed to select the specific one that is deemed to be of physical
relevance. In the case of quantum field theory in a curved space-time
$(\M,\g)$, there is no universal way of doing this unless $\g$ admits
a time-like Killing vector. The analogue in the histories formalism
arises when one tries to construct projection operators like
$\op{P}(f,I)$. Specifically, there will exist infinitely many
unitarily inequivalent representations of the lattice structure of
$\UP$ formed from a particular HPO-theory.

\section{Conclusions}
We have argued that the collection $\UP$ of all history propositions
in a general history theory can be equipped with a lattice structure
that is similar in some respects to the lattice of propositions in
standard quantum logic. Any quasi-temporal properties of the theory
are coded in the space $\S$ of supports associated with the subspace
$\U$ of history filters. The support space can be quite exotic, as
shown by the cobordism example mentioned briefly in section
\ref{SSec:SigTS}. Of course, a Boolean lattice would correspond to a
history version of a classical theory, and quantum-mechanical
superselection rules would arise in the usual way via the existence of
a non-trivial center for the lattice $\UP$.

    We have also seen how an `HPO-theory' exists for standard quantum
theory in which every history proposition can be represented as a
projection operator on a certain Hilbert space. This provides valuable
clues about the possible lattice structure on $\UP$ in the general
case and opens the possibility for novel
\footnote{For example, the lattice structure on $\UP$ can be used to
provide a notion of two history propositions {\em commuting\/}
\cite{BC81}.}
concepts. It also suggests that useful quantum history theories can
be obtained using the lattice of projection operators $P({\cal V})$ on some
Hilbert space $\cal V$ as a model for $\UP$, with a support space $\S$
attached in some way.

    It is clear that much work remains to be done to develop these ideas
into a fully effective tool. Some of the major topics are as follows:

\noindent
{\em The structure of the space $\cal D$ of decoherence
functionals\/}.\ \
\noindent
Very little has been said in this paper about the mathematical
structure of the space of decoherence functionals, but this is clearly
of great importance. In standard quantum logic, the analogue of $\cal
D$ is the space $\cal R$ of states/probability measures on the lattice
$\L$ of single-time propositions. Much has been written about how the
structures of $\cal R$ and $\L$ intertwine, and a similar analysis
should be done for $\cal D$ and $\UP$.

    Another intriguing issue concerns the use of complex numbers. The
only number system that plays a basic role in standard quantum logic
is the set of real numbers in which states take their values. Indeed,
one of the original reasons for interest in the quantum logic
programme was a hope that it would clarify the extent to which the
complex number field is, or is not, an essential ingredient in quantum
theory. The conclusion there was that complex numbers are {\em not\/}
essential, but we have forced them on the history formalism by
requiring decoherence functionals to be complex valued. The question
arises therefore of whether the axioms for decoherence functionals can
be changed so that complex numbers do not play such a fundamental
role.

\medskip
\noindent
{\em A history analogue of the canonical commutation relations.\/}\ \
\noindent
The canonical commutation relations play an important role in
standard Hilbert-space based quantum theory. They arise as the Lie
algebra of a group of transformations (the Weyl group) of classical
phase space and provide the special class of classical observables
that can be assigned self-adjoint operator status in such a way that
Poisson brackets go into operator commutators. More generally, any
classical system with a state space that admits a transitive group of
symplectic transformations $G$ can be quantised by looking for
irreducible, unitary representations of $G$; the associated
self-adjoint representation of the Lie algebra $L(G)$ then provides
the preferred class of quantum observables.
\footnote{This is thinking of a canonical group as a route for
quantising a classical system. Another aspect is the {\em uncertainty
relations\/} that are associated with the commutators of the
generators of $L(G)$. This has been discussed in depth recently by
Halliwell and Anderson \cite{AH93,Hal93a} for the case of the
Weyl group.}

    It is of considerable interest to ask what is the history analogue
of a canonical group. In standard quantum theory on a Hilbert space
$\H$ a unitary representation of $G$ on $\H$ induces a $G$ action on
$P(\H)$; irreducibility of the representation is equivalent to the
action on $P(\H)$ having no fixed points. This idea can be generalised
to the space $F({\cal T},P(\H))$ of history filters by studying the
action on this space of the `gauged' canonical group $F({\cal T},G)$
(\cf the discussion in section \ref{SSec:GrpHPO}). In particular, this
suggests the idea of a `preferred class' of history filters in which
the constituent single-time projection operators all belong to the
spectral resolutions of the Lie algebra of $G$. A detailed study of
this example could lead  to a group-theoretic approach to the actual
{\em construction\/} of general history theories. Of course, the
notion of a {\em symmetry\/} group will also have a natural
translation into history terms.

\medskip
\noindent
{\em Topological aspects of quantum theory in a history formalism.}\ \
\noindent
It is intriguing to wonder how the familiar topological properties of
standard quantum theory translate into the history formalism, and
whether any of them continue to be of significance in a more general
history theory. For example, in the quantum version of a classical
system with configuration space $Q$, different flat, complex vector
bundles over $Q$ lead to inequivalent quantisations. Such bundles are
classified by the group Hom$(\pi_1(Q),U(n))$ of $U(n)$-valued
homomorphisms of the fundamental group $\pi_1(Q)$ of $Q$. How is this
structure reflected in the structure of the space of history
propositions? Similarly, how does the famous Berry phase appear in a
history formalism? These questions may, or may not, be related to the
separate issue of whether the spaces $\U$ and $\UP$ should be given
topological structures in their own right and, if so, what role this
plays in the general theory.

\medskip\noindent
{\em The use of alternative algebraic structures.}\ \
\noindent
The discussion in this paper has been within the context of quantum
logic,\ie the major mathematical tools are to be drawn from the theory
of non-distributive lattices. However, in conventional quantum theory
other types of algebraic structure have been employed to great effect,
and it is important to see if these can be adapted to the general
history formalism. A relatively mild example would be to replace the
use of projection operators with the more general
positive-operator-valued measures that have been much used in recent
discussions of the measurement problem in quantum theory \cite{BLM91}.
A more striking challenge would be to find a $C^*$-algebra analogue of
the general axioms presented in this paper.

\bigskip\noindent
{\Large\bf Acknowledgements}

\noindent
I am extremely grateful to Jonathan Halliwell for numerous discussions
on the consistent history formalism and for a careful reading of a
draft version of this paper. For the latter task I also extend warm
thanks to Arley Anderson and to Jim Hartle.


\begin{thebibliography}{10}
\bibitem{AH93}
A.~Anderson and J.J. Halliwell.
\newblock An information-theoretic measure of uncertainty due to
quantum and thermal fluctuations.
\newblock 1993.
\newblock Imperial College preprint IC 92-93/25.

\bibitem{BC81}
E.G. Beltrametti and G.~Cassinelli.
\newblock {\em The Logic of Quantum Mechanics}.
\newblock Addison-Wesley, London, 1981.

\bibitem{Ble91}
M.~Blencowe.
\newblock The consistent histories interpretation of quantum fields
in curved spacetime.
\newblock {\em Ann. Phys. (NY)}, 211:87--111, 1991.

\bibitem{BLM91}
P.~Busch, P.J. Lahti, and P.~Mittelstaedt.
\newblock {\em The Quantum Theory of Measurement}.
\newblock Springer-Verlag, London, 1991.

\bibitem{DH92} H.F. Dowker and J.J. Halliwell.
\newblock The quantum mechanics of history: {T}he decoherence
functional in quantum mechanics.
\newblock {\em Phys. Rev.}, D46:1580--1609, 1992.

\bibitem{GH90c}
M.~Gell-{M}ann and J.~Hartle.
\newblock Alternative decohering histories in quantum mechanics.
\newblock In K.K. Phua and Y.~Yamaguchi, editors, {\em Proceedings of
the 25th International Conference on High Energy Physics, Singapore,
August, 2--8, 1990}, Singapore, 1990. World Scientific.

\bibitem{GH90a}
M.~Gell-{M}ann and J.~Hartle.
\newblock Quantum mechanics in the light of quantum cosmology.
\newblock In S.~Kobayashi, H.~Ezawa, Y.~Murayama, and S.~Nomura,
editors, {\em Proceedings of the Third International Symposium on the
Foundations of Quantum Mechanics in the Light of New Technology},
pages 321--343. Physical Society of Japan, Tokyo, 1990.

\bibitem{GH90b}
M.~Gell-{M}ann and J.~Hartle.
\newblock Quantum mechanics in the light of quantum cosmology.
\newblock In W.~Zurek, editor, {\em Complexity, Entropy and the Physics of
Information, SFI Studies in the Science of Complexity, {Vol. VIII}}, pages
425--458. Addison-Wesley, Reading, 1990.

\bibitem{GH92}
M.~Gell-{M}ann and J.~Hartle.
\newblock Classical equations for quantum systems.
\newblock 1992.
\newblock UCSB preprint UCSBTH-91-15.

\bibitem{Gle57}
A.M. Gleason.
\newblock Measures on the closed subspaces of a {H}ilbert space.
\newblock {\em Journal of Mathematics and Mechanics}, pages 885--893, 1957.

\bibitem{Gri84}
R.B. Griffiths.
\newblock Consistent histories and the interpretation of quantum mechanics.
\newblock {\em J. Stat. Phys.}, 36:219--272, 1984.

\bibitem{Gui72}
A.~Guichardet.
\newblock {\em Symmetric Hilbert Spaces and Related Topics}.
\newblock Springer-Verlag, New York, 1972.

\bibitem{Hal93a}
J.J. Halliwell.
\newblock Quantum-mechanical histories and the uncertainty principle: {I.
 Information-theoretic} inequalities.
\newblock 1993.
\newblock grqc 9304039.

\bibitem{Har91a}
J.~Hartle.
\newblock The quantum mechanics of cosmology.
\newblock In S.~Coleman, P.~Hartle, T.~Piran, and S.~Weinberg, editors, {\em
  Quantum Cosmology and Baby Universes}. World Scientific, Singapore, 1991.

\bibitem{Har91b}
J.~Hartle.
\newblock Spacetime grainings in nonrelativistic quantum mechanics.
\newblock {\em Phys. Rev.}, D44:3173--3195, 1991.

\bibitem{Har93a}
J.~Hartle.
\newblock Spacetime quantum mechanics and the quantum mechanics of spacetime.
\newblock In {\em Proceedings on the 1992 Les Houches School, Gravitation and
  Quantisation}. 1993.

\bibitem{Hig71}
P.J. Higgins.
\newblock {\em Categories and Groupoids}.
\newblock Van Nostrand, London, 1971.

\bibitem{Ish92}
C.J. Isham.
\newblock Canonical quantum gravity and the problem of time.
\newblock In {\em Proceedings of the NATO Advanced Study Institute, Salamanca,
  June 1992}. Kluwer Academic Publishers, London, 1993.

\bibitem{Jau73}
J.M. Jauch.
\newblock {\em Foundations of Quantum Mechanics}.
\newblock Addison-Wesley, London, 1973.

\bibitem{Kuc92a}
K.~Kucha\v{r}.
\newblock Time and interpretations of quantum gravity.
\newblock In {\em Proceedings of the 4th Canadian Conference on General
  Relativity and Relativistic Astrophysics}. World Scientific, Singapore, 1992.

\bibitem{Mac87}
K.~Mackenzie.
\newblock {\em Lie Groupoids and Lie Algebroids in Differential Geometry}.
\newblock Cambridge University Press, Cambridge, 1987.

\bibitem{Mac63}
G.W. Mackey.
\newblock {\em The Mathematical Foundations of Quantum Mechanics}.
\newblock W.A.~Benjamin, New York, 1963.

\bibitem{Mil65}
J.~Milnor.
\newblock {\em Lectures on the h-Cobordism Theorem}.
\newblock Princeton University Press, Princeton, 1965.

\bibitem{Mit77}
P.~Mittelstaedt.
\newblock Time dependent propositions and quantum logic.
\newblock {\em Jour.~{P}hil.~{L}ogic}, 6:463--472, 1977.

\bibitem{Mit78}
P.~Mittelstaedt.
\newblock {\em Quantum Logic}.
\newblock D.~Reidel, Holland, 1978.

\bibitem{Mit83b}
P.~Mittelstaedt.
\newblock Analysis of the {EPR}-experiment by relativistic quantum logic.
\newblock In {\em Proceedings of the International Symposium on the Foundations
  of Quantum Mechanics}, pages 251--255. 1983.

\bibitem{Mit83a}
P.~Mittelstaedt.
\newblock Relativistic quantum logic.
\newblock {\em Int. J. Theor. Phys.}, 22:293--314, 1983.

\bibitem{Mit85}
P.~Mittelstaedt.
\newblock {EPR-Paradox}, quantum logic and relativity.
\newblock In P.~Lahti and P.~Mittelstaedt, editors, {\em Symposium on the
  Foundations of Modern Physics}, pages 171--186. World Scientific, Singapore,
  1985.

\bibitem{MS83}
P.~Mittelstaedt and E.W. Stachow.
\newblock Analysis of the {EPR}-experiment by relativistic quantum logic.
\newblock {\em Int. J. Theor. Phys.}, 22:517--540, 1983.

\bibitem{Omn88a}
R.~Omn\`es.
\newblock Logical reformulation of quantum mechanics. {I.} {F}oundations.
\newblock {\em J. Stat. Phys.}, 53:893--932, 1988.

\bibitem{Omn88b}
R.~Omn\`es.
\newblock Logical reformulation of quantum mechanics. {II.} {I}nterferences and
  the {E}instein-{P}odolsky-{R}osen experiment.
\newblock {\em J. Stat. Phys.}, 53:933--955, 1988.

\bibitem{Omn88c}
R.~Omn\`es.
\newblock Logical reformulation of quantum mechanics. {III.} {C}lassical limit
  and irreversibility.
\newblock {\em J. Stat. Phys.}, 53:957--975, 1988.

\bibitem{Omn89}
R.~Omn\`es.
\newblock Logical reformulation of quantum mechanics. {III.} {P}rojectors in
  semiclassical physics.
\newblock {\em J. Stat. Phys.}, 57:357--382, 1989.

\bibitem{Omn90}
R.~Omn\`es.
\newblock From {H}ilbert space to common sense: {A} synthesis of recent
  progress in the interpretation of quantum mechanics.
\newblock {\em Ann. Phys. (NY)}, 201:354--447, 1990.

\bibitem{Omn92}
R.~Omn\`es.
\newblock Consistent interpretations of quantum mechanics.
\newblock {\em Rev. Mod. Phys.}, 64:339--382, 1992.

\bibitem{Pir76}
C.~Piron.
\newblock {\em Foundations of Quantum Physics}.
\newblock W.A.~Benjamin, London, 1976.

\bibitem{PP91}
P.~Pt\'ak and S.~Pulmannov\'a.
\newblock {\em Orthomodular Structures as Quantum Logics}.
\newblock Kluwer Academic Publishers, London, 1991.

\bibitem{Sor93a}
R.D. Sorkin.
\newblock Impossible measurements on quantum fields.
\newblock In B.S. Hu and T.A. Jacobson, editors, {\em Directions in General
  Relativity, Vol. II: a Collection of Essays in honour of Dieter Brill's
  Sixtieth Birthday}. Cambridge University Press, Cambridge, 1993.

\bibitem{Sta80}
{E.-W.} Stachow.
\newblock Logical foundations of quantum mechanics.
\newblock {\em Int. J. Theor. Phys.}, 19:251--304, 1980.

\bibitem{Sta81}
{E.-W.} Stachow.
\newblock Sequential quantum logic.
\newblock In E.G. Beltrametti and B.C. {van Fraassen}, editors, {\em Current
  Issues in Quantum Logic}, pages 173--191. Plenum Press, New York, 1981.

\bibitem{VNB36}
J.~{v}on Neumann and G.~Birkhoff.
\newblock The logic of quantum mechanics.
\newblock {\em Annals of Mathematics}, 37:823--843, 1936.

\end{thebibliography}
\end{document}